\documentclass[a4paper,fleqn,usenatbib,useAMS]{mn3e}

\usepackage[T1]{fontenc}
\usepackage{ae,aecompl}

\RequirePackage[colorlinks,citecolor=blue,urlcolor=blue]{hyperref}

\usepackage{amsmath, amssymb}
\usepackage{algpseudocode}
\usepackage{algorithm}
\usepackage{graphicx}
\usepackage{subfigure}
\usepackage{lineno}
\usepackage{afterpage}
\usepackage{comment}
\usepackage{enumitem}

\let\hat\widehat

\newcommand\E{\mathbb{E}}

\newcommand\norm[1]{\|#1\|}
\newcommand\Step{{\sf Step }}
\newcommand\FCov{{\sf FCov}}
\newcommand\Cov{{\sf Cov}}

\usepackage{pbox}

\title[Cosmic Web Reconstruction through Density Ridges]{Cosmic Web Reconstruction through 
Density Ridges: Method and Algorithm}
\author[Yen-Chi Chen et al.]{Yen-Chi Chen,$^{1,3}$\thanks{E-mail:
yenchic@andrew.cmu.edu}
Shirley Ho,$^{2,3}$
Peter E. Freeman,$^{1,3}$ \newauthor
Christopher R. Genovese,$^{1,3}$
Larry Wasserman$^{1,3}$
\\
$^{1}$Department of Statistics, Carnegie Mellon University, Pittsburgh, PA 15213, USA\\
$^{2}$Department of Physics, Carnegie Mellon University, Pittsburgh, PA 15213, USA\\
$^{3}$McWilliams Center for Cosmology, Department of Physics, Carnegie Mellon University, Pittsburgh, PA 15213, USA}
\begin{document}


\pagerange{\pageref{firstpage}--\pageref{lastpage}} \pubyear{2015}

\maketitle

\label{firstpage}

\begin{abstract}

The detection and characterization of filamentary structures in the cosmic
web allows cosmologists to constrain parameters that dictate 
the evolution of the Universe.
While many filament estimators have been proposed, they generally lack
estimates of uncertainty, reducing their inferential power. In this
paper, we demonstrate how one may apply the Subspace Constrained Mean Shift (SCMS) algorithm
(\citealt{Ozertem2011}, \citealt{2012arXiv1212.5156G}) 
to uncover filamentary structure in galaxy
data. The SCMS algorithm is a gradient ascent method that models filaments
as density ridges, one-dimensional smooth curves that trace high-density 
regions within the point cloud. We also demonstrate how augmenting the SCMS
algorithm with bootstrap-based methods of uncertainty estimation allows one
to place uncertainty bands around putative filaments. 
We apply the SCMS first to the dataset generated from the Voronoi model.
The density ridges show strong agreement with the filaments from Voronoi method.
We then apply the SCMS 
method datasets
sampled from a P3M N-body simulation, with galaxy number
densities consistent with SDSS and {\it WFIRST-AFTA},
and to LOWZ and CMASS data from the Baryon Oscillation Spectroscopic Survey
(BOSS). To further assess the efficacy of SCMS, we compare the relative
locations of BOSS filaments with galaxy clusters in the redMaPPer catalog,
and find that redMaPPer clusters are significantly closer (with
p-values $< 10^{-9}$) to SCMS-detected
filaments than to randomly selected galaxies.
\end{abstract}


\section{Introduction}

Observations of the local universe made over the last 
four decades show that on megaparsec scales,
matter is distributed in web-like structures$-$clusters, filaments,
sheets, and voids$-$that arise naturally from the non-linear evolution
of initially small density fluctuations~\citep{Peebles1980,Bond1996,
Jenkins1998,Colberg2005,2005Natur.435..629S,2006MNRAS.370..656D}.
Of particular interest to us are the filaments, one-dimensional structures
that connect galaxy clusters and form at the boundaries of empty voids.
Filaments are of interest for several reasons.
The detection and characterization of filaments at
a range of redshifts provides a means by which cosmologists can constrain
theories of the universe's evolution
\citep{Bond1996,2009ApJ...706..747Z,2013ApJ...779..160Z}.
Filaments also influence 
the shape, angular momentum, and peculiar velocities
of dark matter haloes 
\citep{2007MNRAS.381...41H,2007MNRAS.375..489H,2008MNRAS.389.1127P,
2009MNRAS.398.1742H,2009ApJ...706..747Z,Jones2010,
2013ApJ...779..160Z,ForeroRomero2014},
as well as the intrinsic alignments and luminosities of nearby galaxies
\citep{2015ApJ...800..112G,2014arXiv1402.3302C,codis2014}.

As the review of \cite{Cautun2014} amply demonstrates,
the detection of filamentary structure is a non-trivial problem for which
many solutions have been proposed. These solutions include methods that 
examine the Hessian matrix of the galaxy density field,
such as the Multiscale Morphology Filter 
(MMF; \citealt{AragonCalvo2007,2010MNRAS.408.2163A}) and
NEXUS and NEXUS+ \citep{Cautun2013}, as well as segmentation-based
methods, such as the Candy model \citep{Stoica2007,2005A&A...434..423S},
the skeleton \citep{2006MNRAS.366.1201N}, 
the Spine method \citep{2010ApJ...723..364A},
and DisPerSE models \citep{2011MNRAS.414..350S}, and
the path density method~\citep{Genovese2009}.
While all of these methods provide estimates of filamentary structure,
none provide an assessment of estimator uncertainty. The fact that
filament estimates are random sets presents a significant challenge to the
construction of valid uncertainty measures \citep{Molchanov2005}. 
 
In this paper, we introduce a new method for filament detection
based on the Subspace Constrained Mean Shift (SCMS)
algorithm of \cite{Ozertem2011}.
The statistical properties of SCMS
were studied in
\cite{2012arXiv1212.5156G}.
The mathematical properties of density ridges and the statistical consistency
of SCMS are discussed in~\cite{Eberly1996,2012arXiv1212.5156G}, and
\cite{2014arXiv1406.5663C}, respectively, while \cite{2014arXiv1406.5663C}
introduce an uncertainty measure to the ridge formalism that allows
one to quantitatively assess, in the context of the current paper, 
putative cosmic filaments.

In {\S}2, we describe the SCMS algorithm and the methods we use to
assess the uncertainty of its filament estimates. 
In {\S}3, we apply SCMS, first to a P3M N-body
simulation output \citep{2015arXiv150702685T}, and then to low-redshift
(0.235 $\leq z \leq$ 0.240) and high-redshift (0.530 $\leq z \leq$ 0.535)
data collected by the Baryon Oscillation Spectroscopic Survey (BOSS), which was
released as part of SDSS Data Release 11. We also demonstrate the 
consistency between filaments detected by SCMS and galaxy clusters
listed in the redMaPPer catalog. In {\S}4 we summarize our results and
offer possible avenues for future methodological development.
In Appendix A we provide further detail on how to optimally select
values for the tuning parameters of the SCMS algorithm, while in
Appendix B we apply the algorithm to labeled simulated data generated
via the Voronoi model of \citep{1994A&A...283..361V} to show that it
preferentially detects structures labeled as filaments. In a second
paper, we will provide a full catalogue of filaments detected in 
SDSS data.

\begin{figure*}
\centering
\subfigure[]
	{
	\includegraphics[scale=0.37]{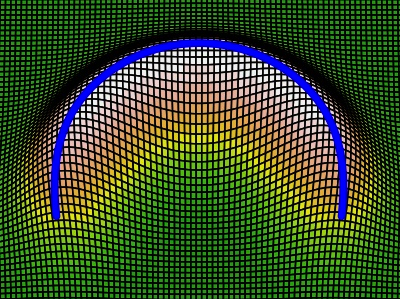}
	}
\subfigure[]
	{
	\includegraphics[scale=0.27]{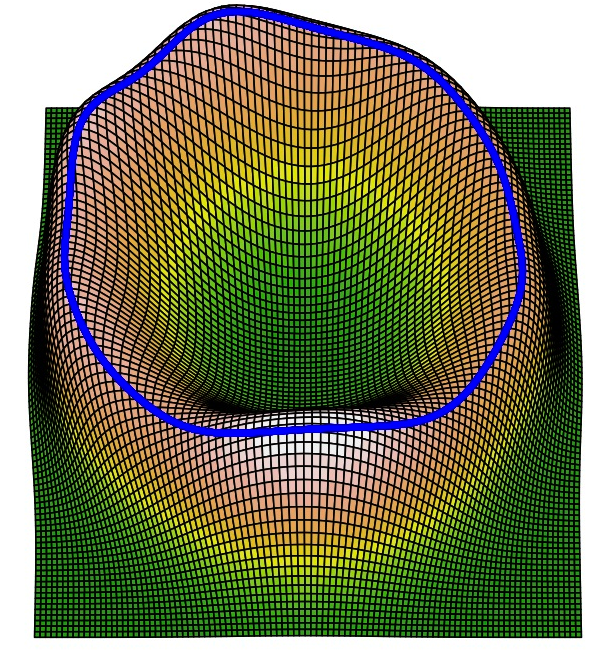}
	}
\caption{Examples of ridges (blue curves) in a smooth function.}
\label{fig::ex}
\end{figure*}

\section{Subspace Constrained Mean Shift: Algorithm}	\label{sec::methods}

\subsection{Density Ridge Formalism} \label{sec::ridges}

Assume that we observe $n$ galaxies with locations 
$X_1,\cdots,X_n$ that are $d-$dimension points;
for data from typical astronomical surveys, 
$d=2$ (if the galaxies are constrained
to a redshift shell) or $d=3$.
We model $X_1,\cdots,X_n$ as random variables sampled 
from an unknown density function $p$. 

Formally, a \emph{density ridge}
\citep{Eberly1996,Ozertem2011,2012arXiv1212.5156G,2014arXiv1406.5663C,2014arXiv1406.1803C}
of $p$ is defined as follows.
Let $g(x) = \nabla p(x)$ and $H(x)$
be the gradient and Hessian, respectively, of $p(x)$
and let $v_1(x),\cdots,v_d(x)$ be the eigenvectors of the Hessian matrix,
with associated eigenvalues 
$\lambda_1(x) \ge \lambda_2(x)\ge \cdots \ge \lambda_d(x)$.
We define
$V(x)$ to be the matrix of all eigenvectors orthogonal to the first,
$[v_2(x),\cdots,v_d(x)]$, and the ridge set $R$ as
\begin{equation} \label{eq::ridge-def}
R \equiv \mbox{Ridge}(p) = \{x: G(x) = 0, \ \lambda_2(x) < 0\} \,,
\end{equation}
where 
\begin{align}
G(x) = V(x)V(x)^T g(x)
\label{eq::PG}
\end{align}
is the projected gradient. 
The fact that ridges have projected a gradient of $0$ (and second eigenvalues being negative) 
means that ridges are local maximums in the subspace
spanned by eigenvectors $v_2(x),\cdots,v_d(x)$.
When $p$ is smooth and the \emph{eigengap}
\begin{align}
\beta(x) = \lambda_1(x) - \lambda_2(x)
\label{eq::eigengap}
\end{align}
is positive, the ridges have the properties of filaments, 
i.e.~smooth curve-like structures with high density (see Figure~\ref{fig::ex}). 
Note that $R$ will include modes
of the density $p$ which, in the context of cosmic filament detection, means 
that $R$ contains both filaments and galaxy clusters.
Also note that density ridges are more general objects 
than the skeleton models proposed in \cite{2006MNRAS.366.1201N, 2008MNRAS.383.1655S}
and the Spine method \citep{2010ApJ...723..364A}.
Essentially, when $d=2,3$, density ridges are the same as skeletons.

Compared with other models, density ridges adapt information from both
gradient and Hessian matrix of density.
In contrast, MMF \citep{AragonCalvo2007,2010MNRAS.408.2163A},
NEXUS and NEXUS+ \citep{Cautun2013}
only use the information of second derivatives (they define filaments
as the regions with $\lambda_2(x)<0$ and $\lambda_1(x)\approx\lambda_2(x)> \lambda_3(x)$).
DisPerSE models \citep{2011MNRAS.414..350S} define filaments as those gradient flows 
that start from saddle points and end up at local maximums, which 
utilize only the first derivatives.

An attractive feature for the density ridge model is that 
the statistical theory for consistently estimating the density ridge
has been well-established 
\citep{2012arXiv1212.5156G,2014arXiv1406.5663C,2014arXiv1406.1803C}.
We also use N-body simulation to verify the convergence 
of density ridges when we subsample different number of galaxies 
(see Section \ref{sec::simulation}).

\begin{figure*}
\centering
\includegraphics[width=2 in]{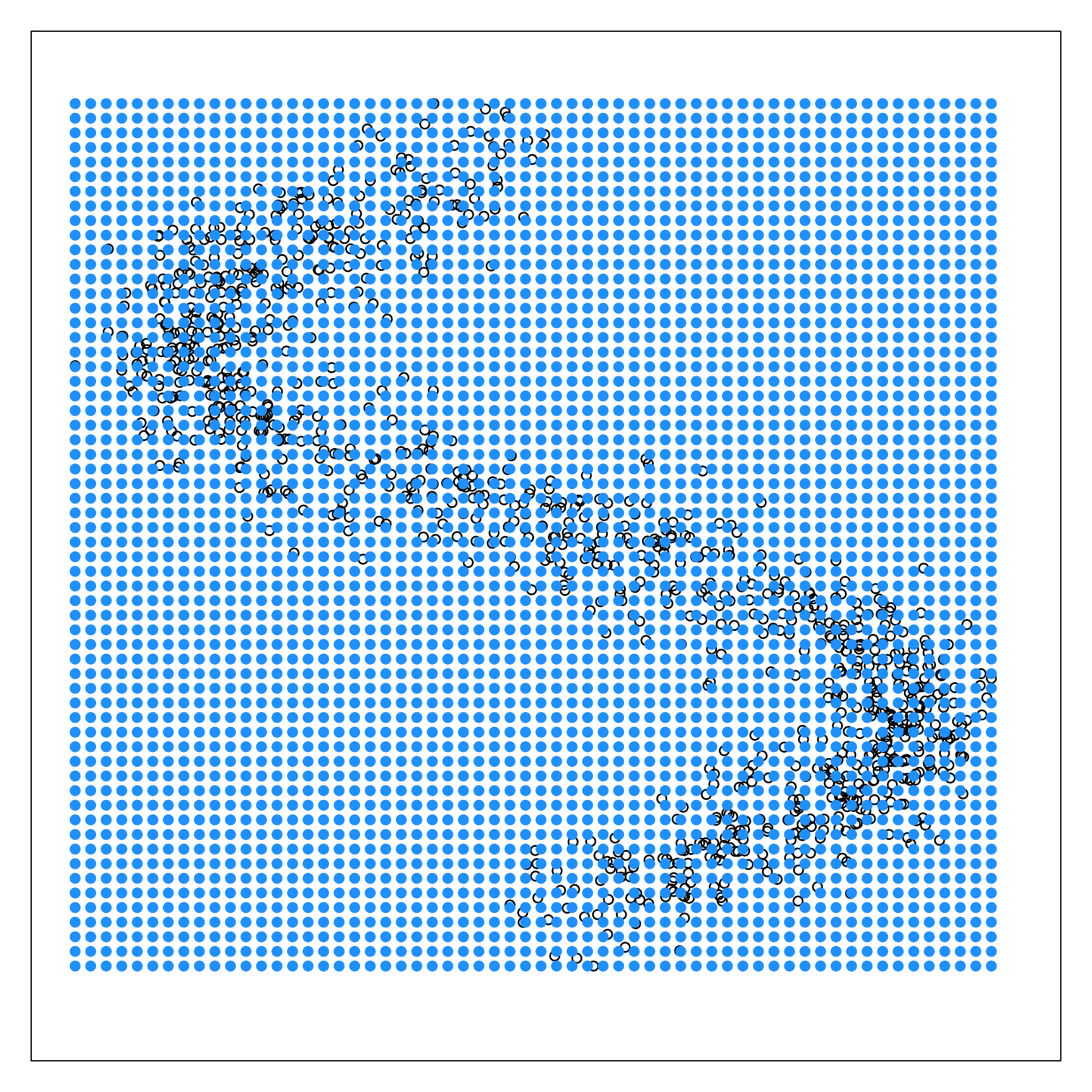}
\includegraphics[width=2 in]{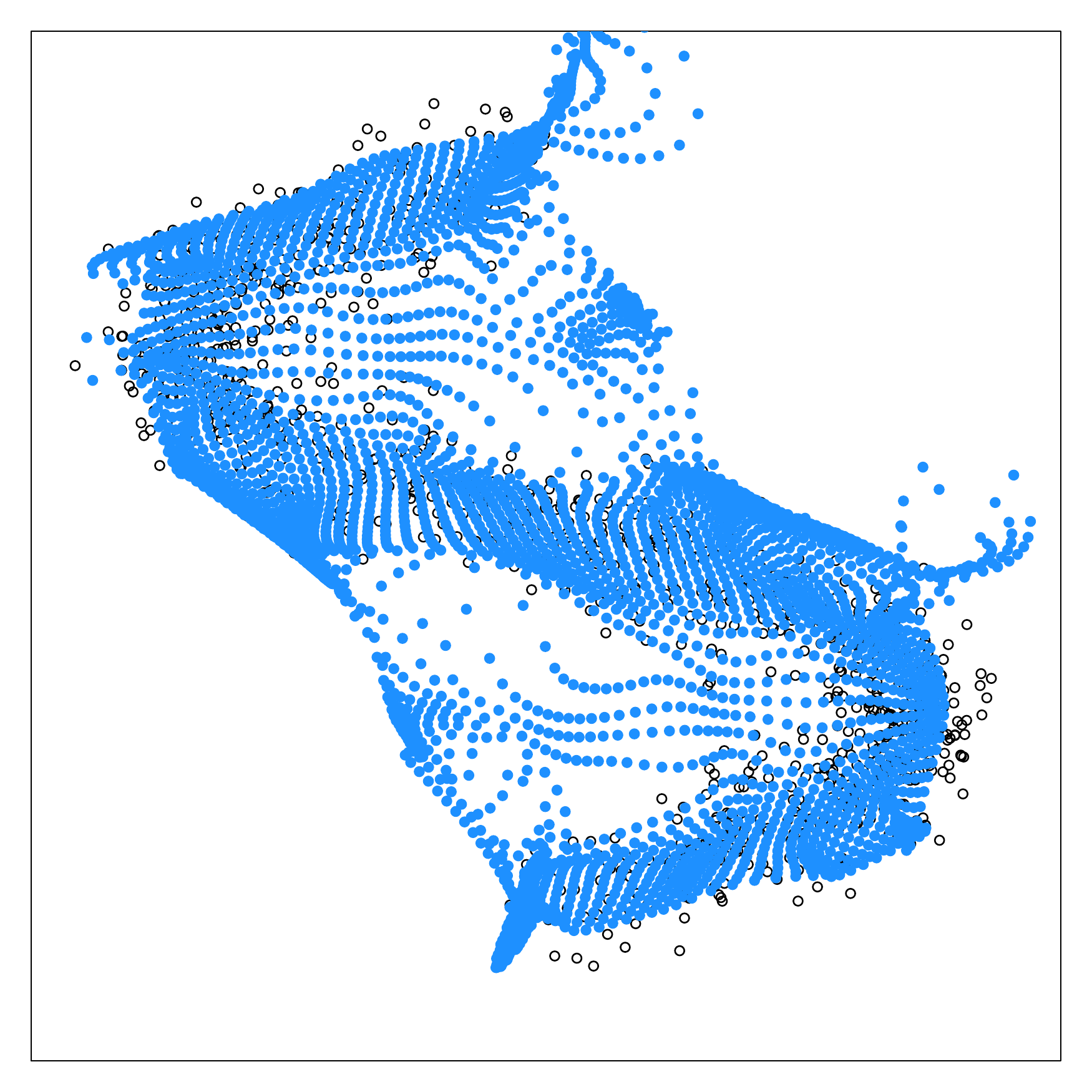}
\includegraphics[width=2 in]{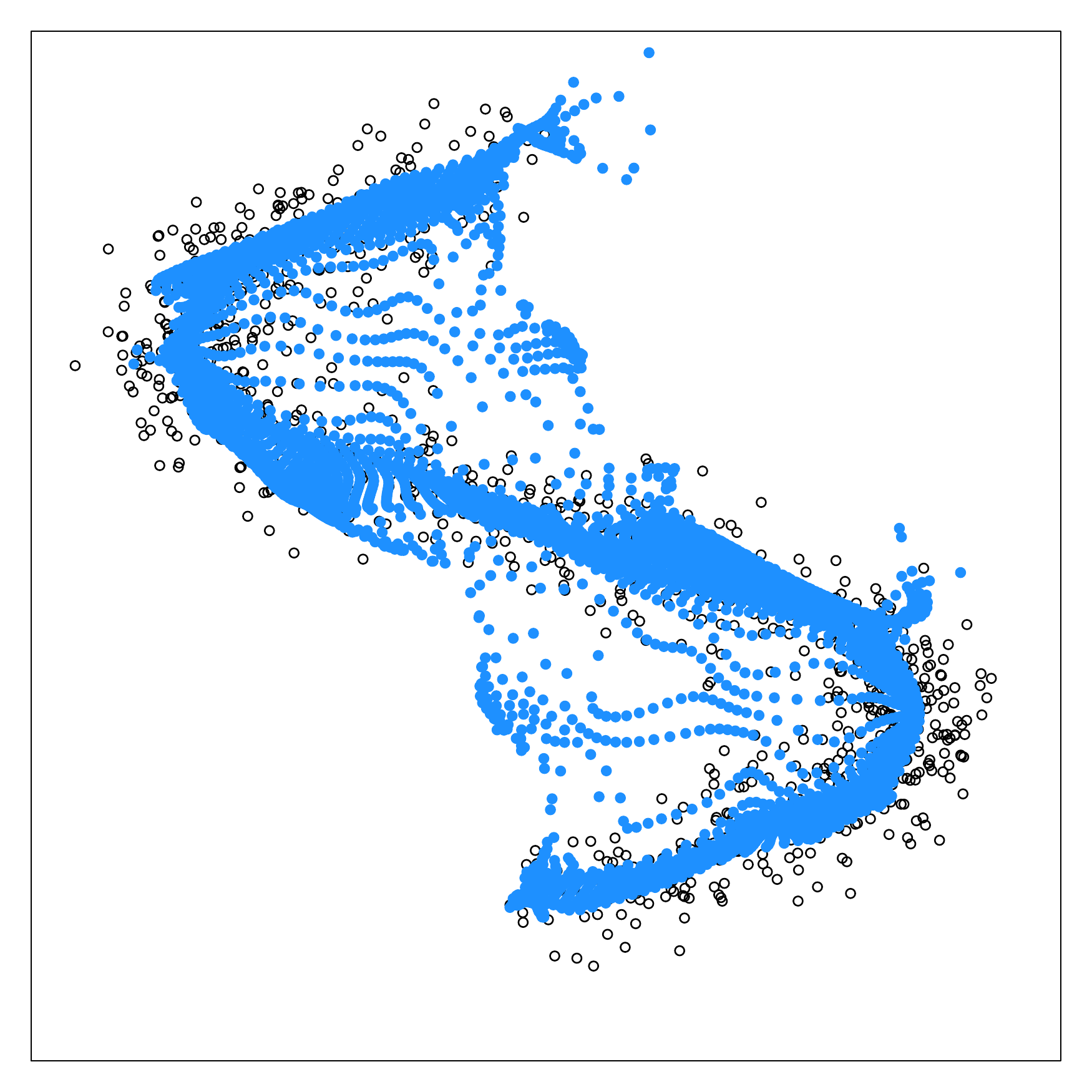}\\
\includegraphics[width=2 in]{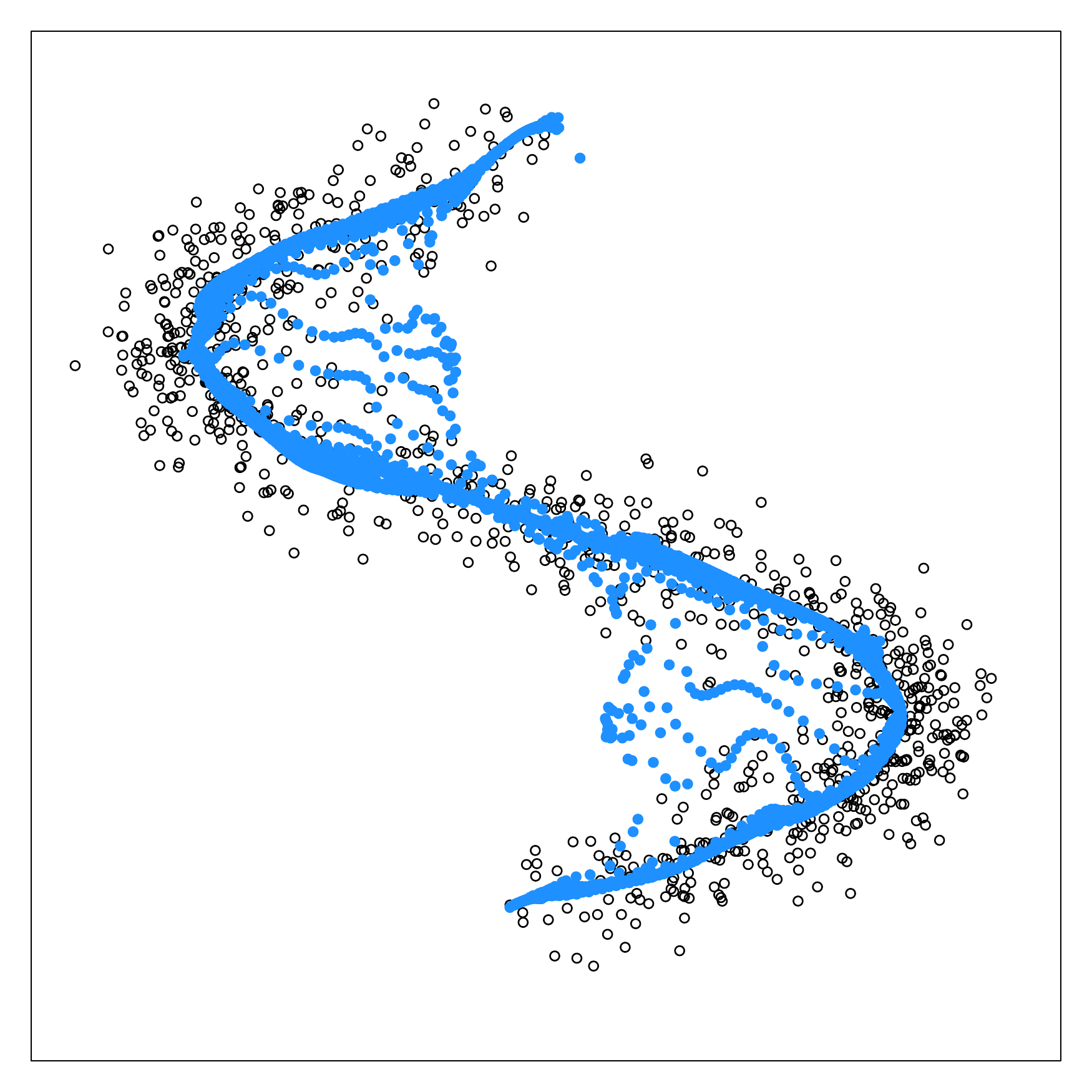}
\includegraphics[width=2 in]{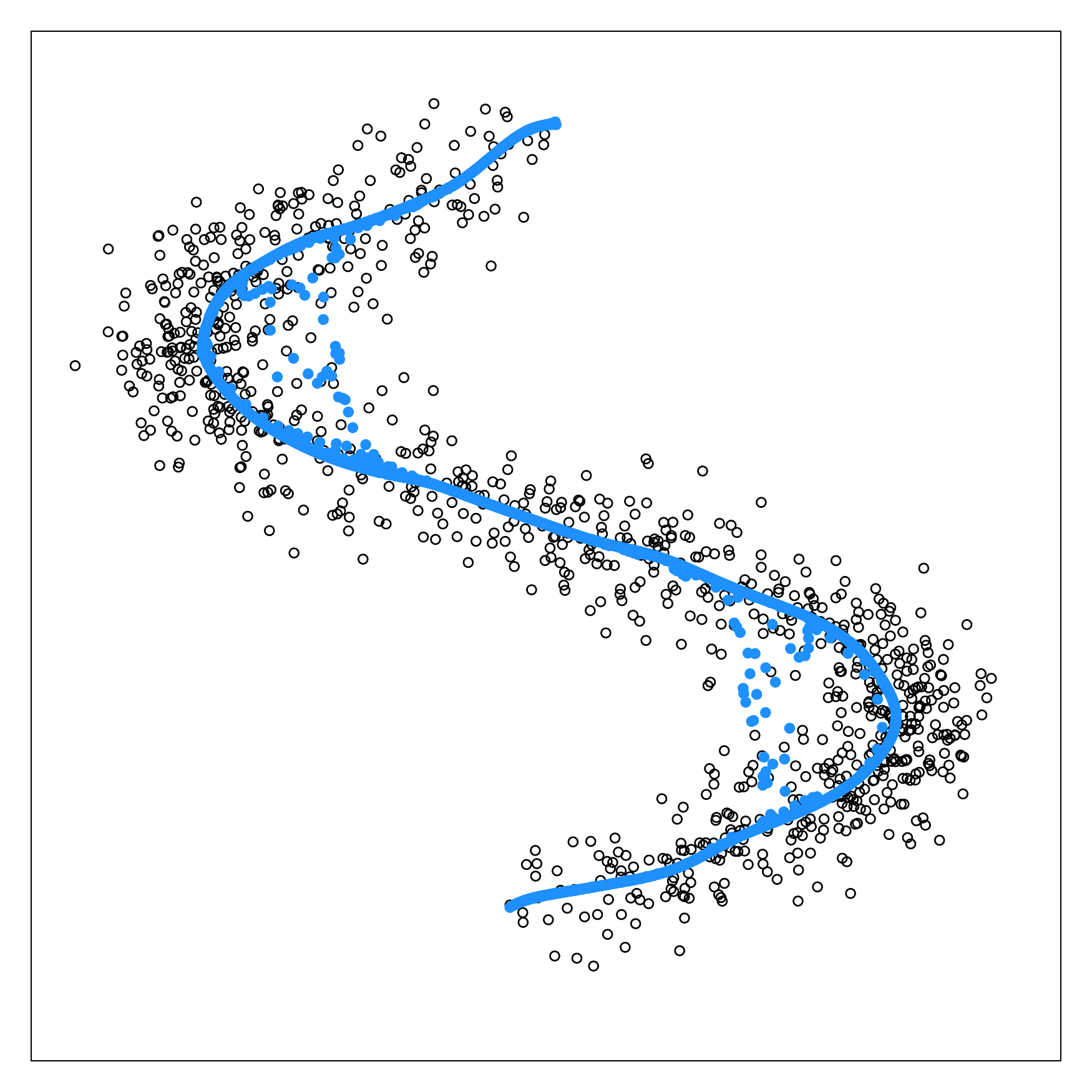}
\includegraphics[width=2 in]{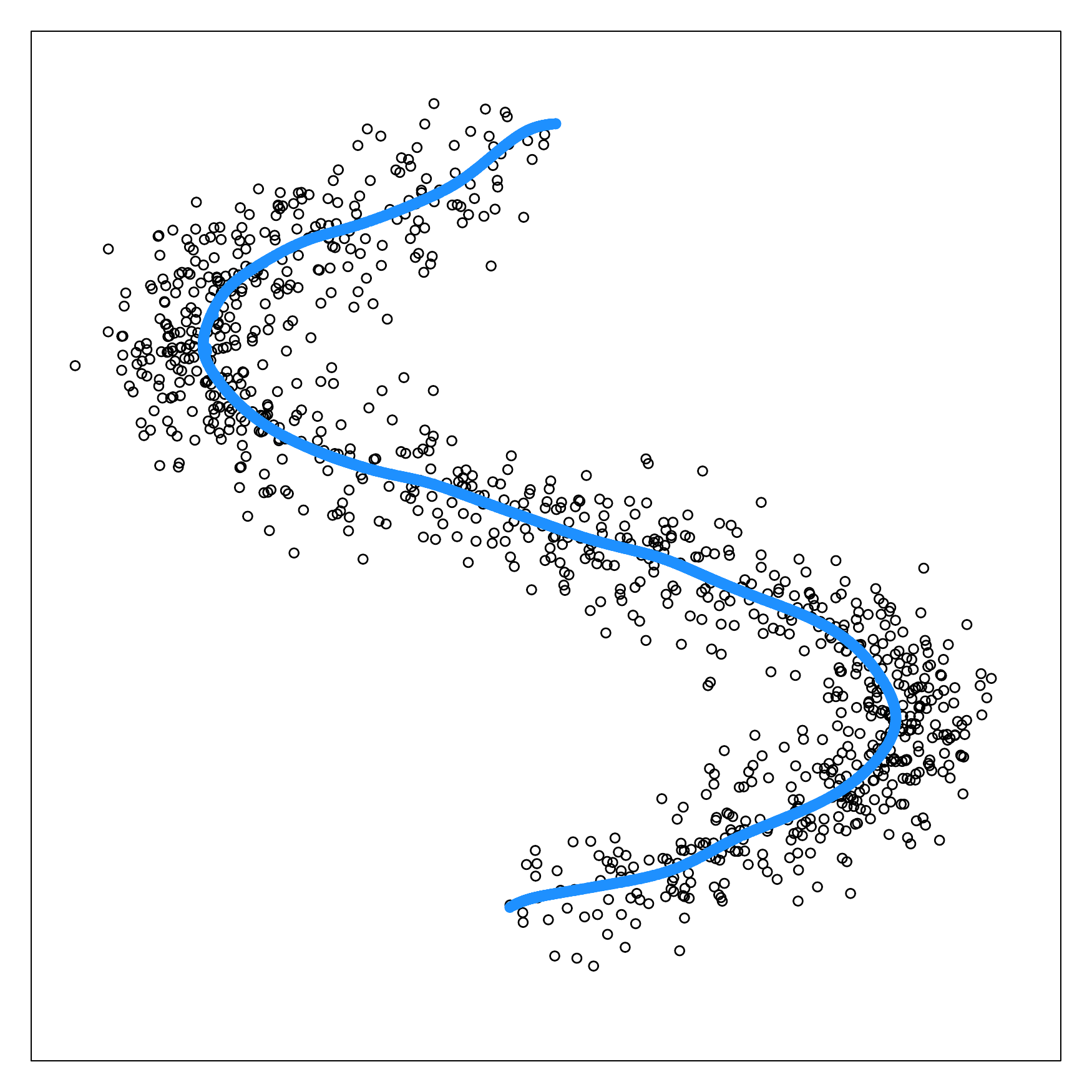}
\caption{Pictoral overview of the SCMS algorithm 
  (Step 4 in Algorithm \ref{alg::FE}).
  Each point in an initially uniform mesh (the blue dots in the top-left panel)
  is moved to the closest density ridge (bottom right).
  The top-middle, top-right, bottom-left, bottom-middle, and bottom-right 
  panels indicate the locations of the mesh points after 1, 2, 4, 8, and 16 
  iterations of the algorithm, respectively.}
\label{fig::SCMS}
\end{figure*}

\subsection{SCMS: Filament Detection}
\label{sec::FE}

\begin{figure*}
\centering
\subfigure[Without thresholding]
	{
	\includegraphics[width=2 in, height= 2 in]{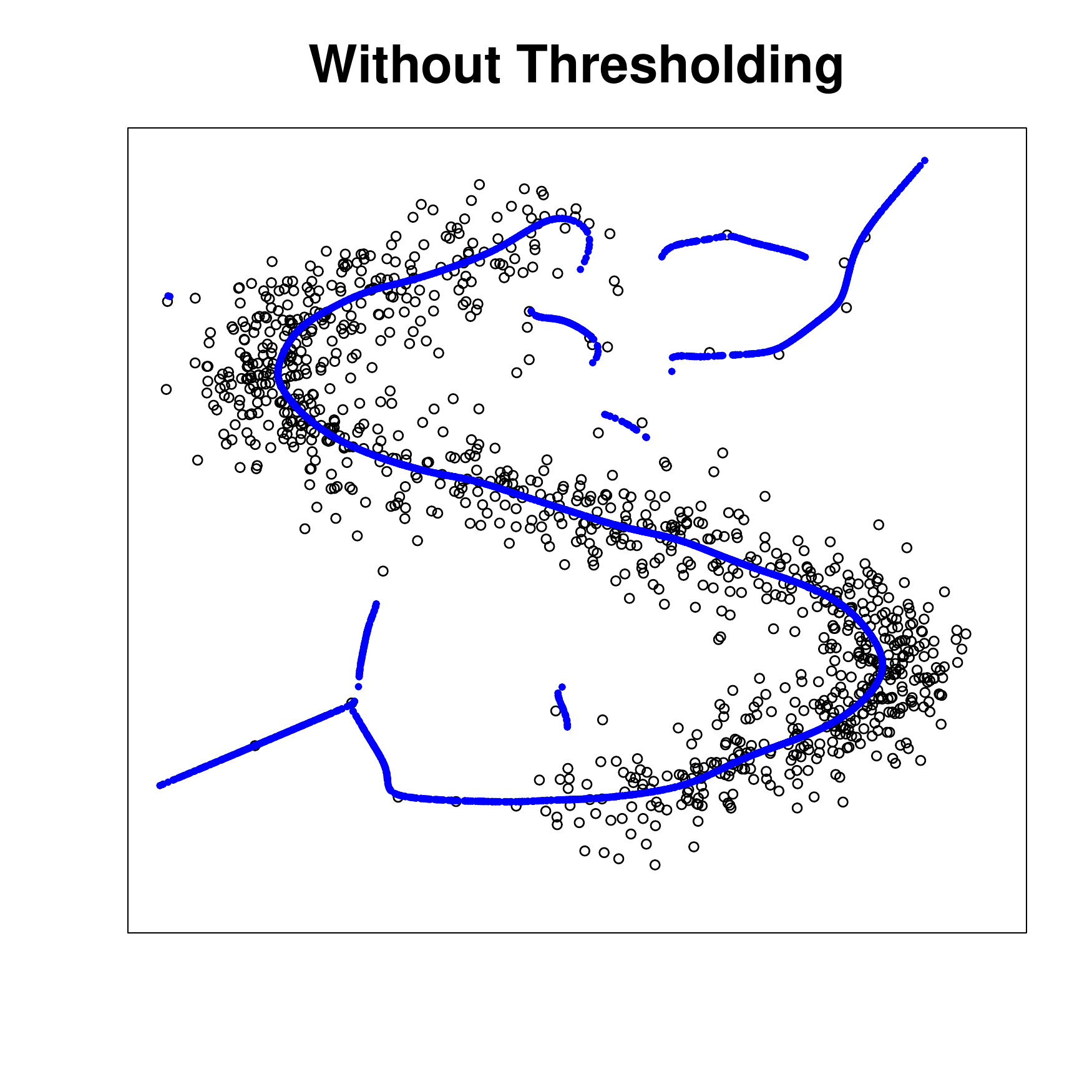}
	}
\subfigure[Thresholding]
	{
	\includegraphics[width=2 in, height= 2 in]{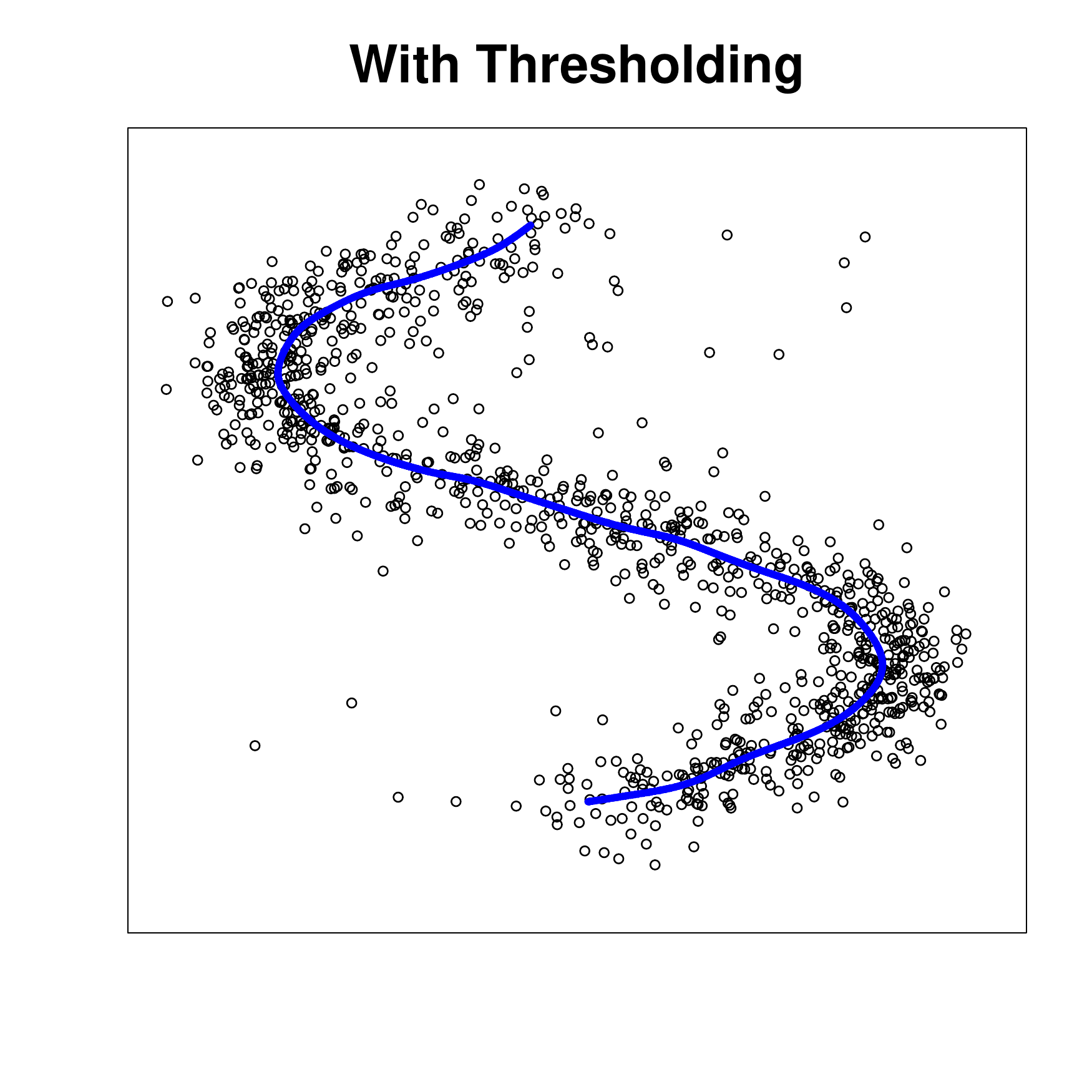}
	}
\caption{An example of the comparison of SCMS with and without noise removal.
This is a simple simulated dataset with clutter noise. 
As can be seen easily, thresholding the density removes problems of clutter noise
.}
\label{fig:SCMSvsSCMS}
\end{figure*}

\begin{figure*}
\centering
\subfigure[]
	{
	\includegraphics[width=2 in, height= 2 in]{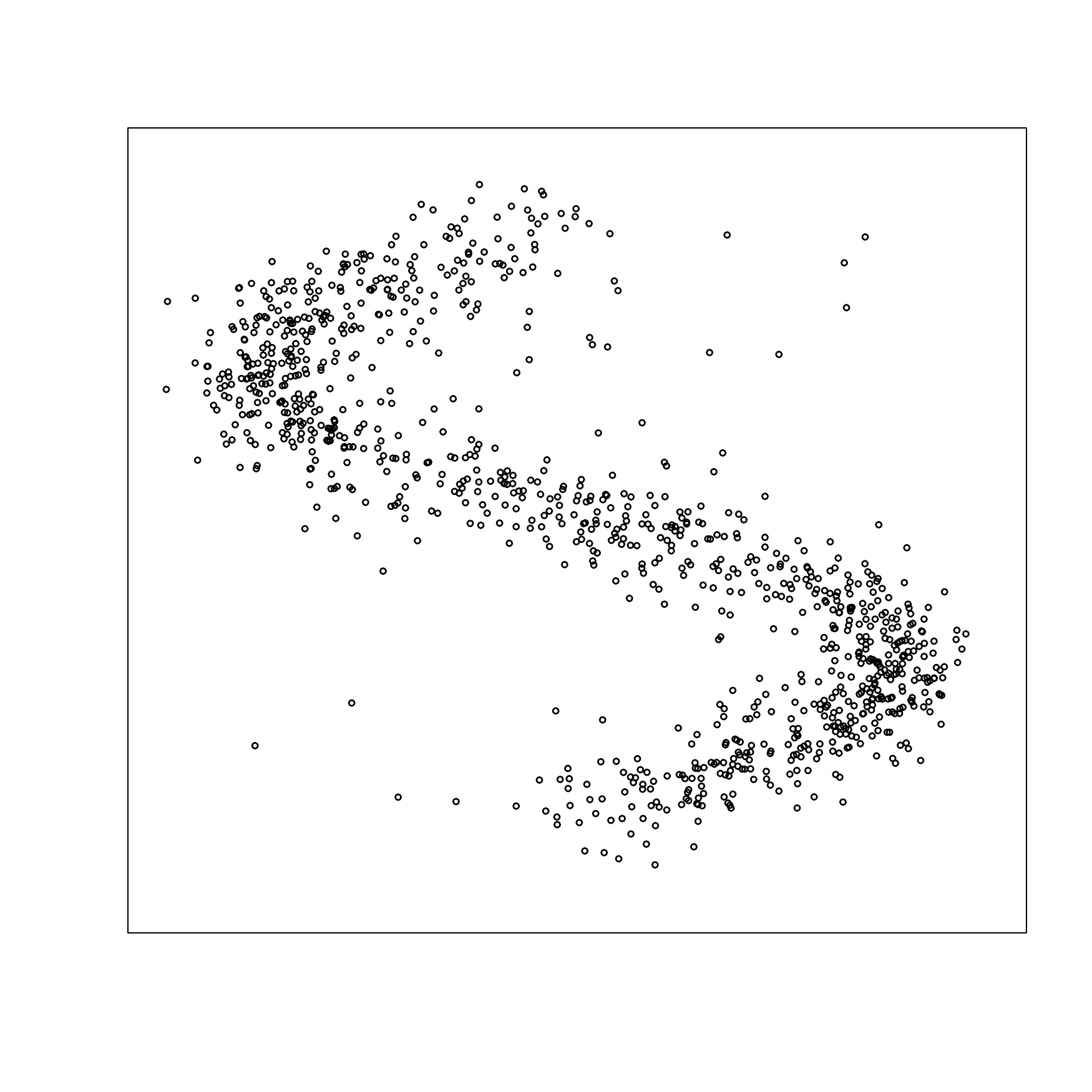}
	}
\subfigure[]
	{
	\includegraphics[width=2 in, height= 2 in]{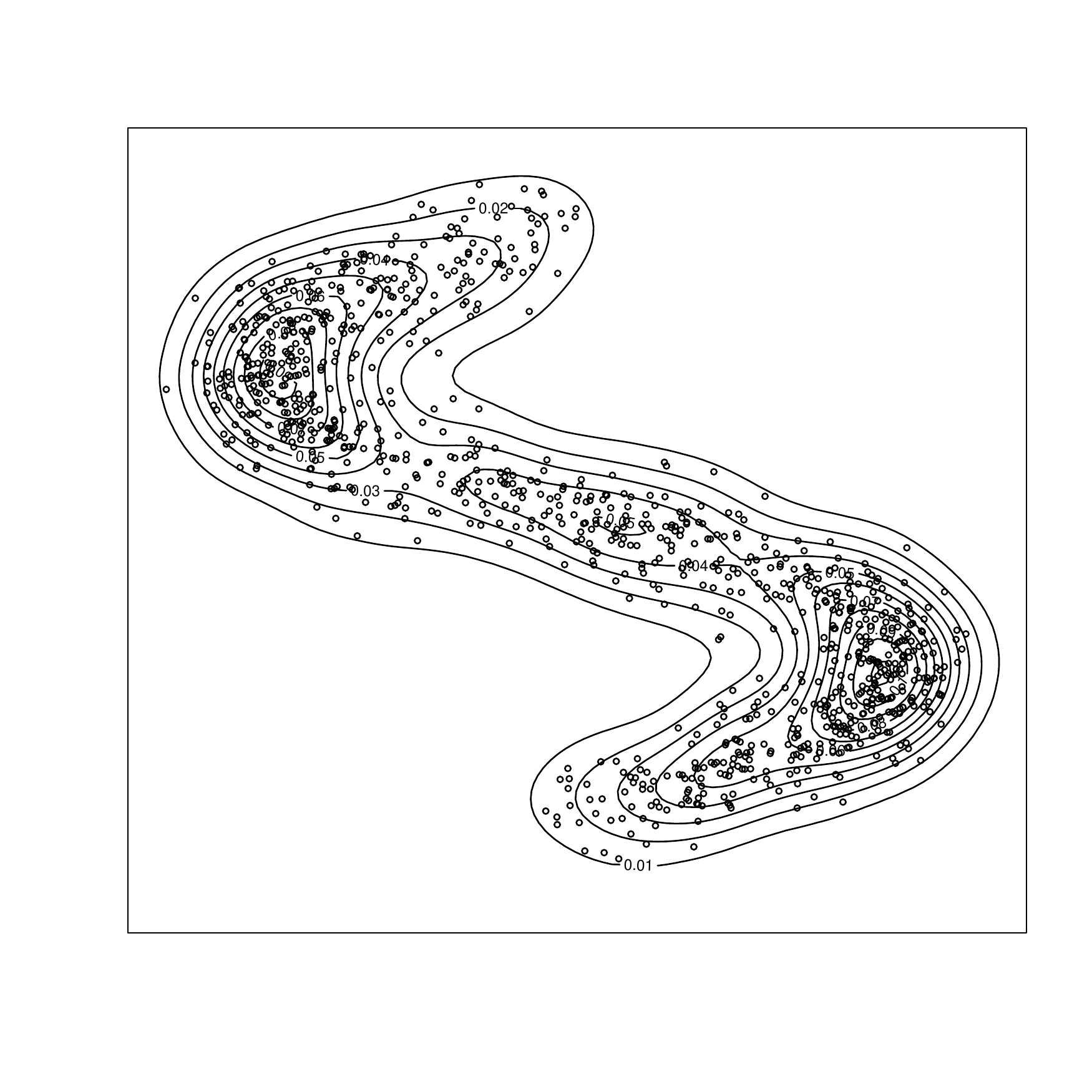}
	}
\subfigure[]
	{
	\includegraphics[width=2 in, height= 2 in]{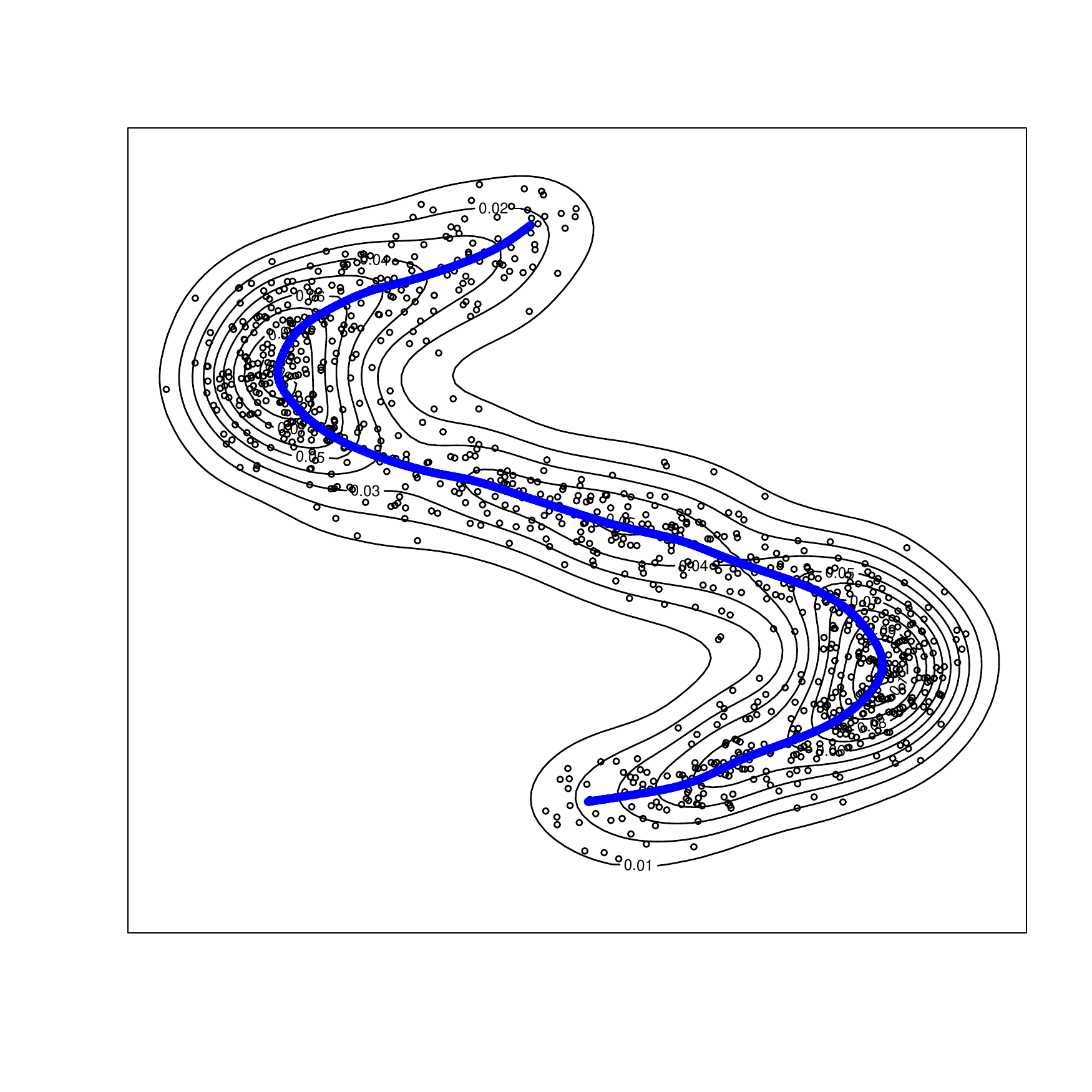}
	}
\caption{An example of the application of SCMS. 
{\bf (a)} The original data. 
{\bf (b)} Contour plot showing the kernel density estimate of the density $p$. 
{\bf (c)} The ridge estimate (blue curve). 
Note that in (b), we remove points where the estimated density is less than a threshold $\tau$.}
\label{fig:RidgeEstimation}
\end{figure*}

The algorithm consists of three 
steps described below and listed in Algorithm \ref{alg::FE}.
The first is to estimate the underlying density function $p(x)$ given
$X_1,\cdots,X_n$, the observed locations of galaxies. We use the
standard kernel density estimator (see e.g.~\citealt{AllNonPar}):
\begin{align}
\hat{p}(x) = \frac{1}{nh^d} \sum_{i=1}^n K\left(\frac{\norm{x-X_i}}{h}\right) \,,
\label{eq::KDE}
\end{align}
where $K(\cdot)$ is the smoothing kernel 
(e.g.~a Gaussian), 
$\norm{x-X_i}$ is the Euclidean distance between the point
$x$ and the $i^{\rm th}$ galaxy location $X_i$, and $h$ is the 
smoothing bandwidth (the selection of which is discussed in
Appendix \ref{sec::parameters}).

In the second step, we denoise by applying a 
threshold to the estimated density function $\hat{p}(x)$
to eliminate the effect that
galaxies in low-probability density regions, i.e.~where
$\hat{p}(x) < \tau$, would have on filament estimation. 
How one selects $\tau$ is also discussed in Appendix 
\ref{sec::parameters}.
The denoising step is not part of the original SCMS algorithm but is important
to increase its statistical power in low-density regions 
(see Figure \ref{fig:SCMSvsSCMS}. 
We note that a thresholding step is included in several filament-detection
algorithms, including those of e.g.~\cite{2006MNRAS.366.1201N} and \cite{2011MNRAS.414..350S}.
%
%

For the final step, given a set of galaxies in high-density regions,
we apply the original version of the SCMS
\citep{Ozertem2011} to detect filamentary structures.
Given a point $x$ on a defined, uniform mesh, 
SCMS moves it according to 
an ``estimated projected gradient" given by
\begin{equation}
\hat{G}(x) = \hat{V}(x)\hat{V}(x)^T \hat{g}(x) \,,
\end{equation}
where $\hat{V}(x),\hat{g}(x)$ are estimates of the 
quantities $V(x),g(x)$ that we define above in Section \ref{sec::ridges}.
One may view this procedure as estimating a ridge set $R$ by applying the 
Ridge operator to $\hat p$:
\begin{align}
\hat{R} = \mbox{Ridge}(\hat{p}).
\end{align}
Essentially, $\hat{R}$ is very similar
to the filaments defined in \cite{2008MNRAS.383.1655S,2010MNRAS.409..156B,2010MNRAS.406..320C}.
Note that a putative filament is, in the context of this algorithm, a set of 
points and not a one-dimensional curve.
In Step 4 of Algorithm \ref{alg::FE},
We further describe how we apply
SCMS.
In Figure \ref{fig::SCMS}, we illustrate the application of SCMS
to uniform mesh of points, and in Figure \ref{fig:SCMSvsSCMS} we 
demonstrate the importance of the thresholding step: the left and
right panels show putative filaments detected without and with thresholding, 
respectively. We observe that thresholding greatly decreases the rate of 
false filament detection.




\begin{algorithm*}
\caption{SCMS (Subspace Constrained Mean Shift)}
\begin{algorithmic}
\State \textbf{Input:} Data $\{ X_1,\cdots,X_n\}$. Smoothing bandwidth $h$. Threshold $\tau$.
\begin{itemize}[leftmargin=1.5cm]
\item[\Step 1.] Compute the density estimator $\hat{p}(x)$ via equation \eqref{eq::KDE}.
\item[\Step 2.] Select a mesh $\mathcal{M}$ of points. By default, we can take $\mathcal{M} = \{X_1,\cdots, X_n\}$.
\item[\Step 3.] Thresholding: remove $m\in\mathcal{M}$ if $\hat{p}(m)<\tau$. Let the remaining mesh points be denoted $\mathcal{M}'$.
\item[\Step 4.] For each $x \in\mathcal{M}'$, perform the following subspace constrained mean shift until convergence:
	\begin{itemize}
\item[\Step 4-1.] For $i=1,\cdots,n$, compute
\begin{align*}
\mu_i = \frac{x-X_i}{h^2},\qquad c_i=K\left(\frac{x-X_i}{h}\right)
\end{align*}
\item[\Step 4-2.] Compute the Hessian matrix
\begin{equation}
H(x)  = \frac{1}{n}\sum_{i=1}^nc_i\left(\mu_i\mu_i^T-\frac{1}{h^2}\mathbf{I}\right).
\label{eq::Hessian}
\end{equation}
\item[\Step 4-3.] Perform spectral decomposition on $H(x)$ and compute $V(x)=\left(v_2(x),\cdots,v_d(x)\right)$,
the eigenvectors corresponding to the smallest $d-1$ eigenvalues.
\item[\Step 4-4.] Update $x\longleftarrow V(x)V(x)^T m(x)+x$ until convergence, where 
\begin{align}
m(x) = \frac{\sum_{i=1}^n c_iX_i}{\sum_{i=1}^nc_i}-x
\label{eq::MS}
\end{align}
is called the mean shift vector.
	\end{itemize}
\end{itemize}
\smallskip
\State \textbf{Output:} The collection of all remaining points.
\end{algorithmic}
\label{alg::FE}
\end{algorithm*}


\subsection{SCMS: Filament Uncertainty Estimation}	\label{sec::LU}

We quantify the uncertainty in the filament estimates produced by SCMS
using the concept of \emph{local uncertainty}~\citep{2014arXiv1406.5663C}.
The local uncertainty in an estimated filament $\hat R$ at a point $x$
on the true filament $R$ is the expected distance between $x$ and
the closest point to $x$ on $\hat R$.  This is denoted by $\rho(x)$
and is given by:
\begin{equation} \label{eq::local-uncertainty}
\rho(x) = \begin{cases}
            \sqrt{ \E \left[d_{\hat{R}}^2(x)\right] } & \mbox{if $x \in R$} \\
            0                       & \mbox{\rm otherwise} 
            \end{cases},
\end{equation}
where $d_{\hat{R}}(x) = \min\{\norm{x-y}: y \in \hat{R}\}$
and the notation $\mathbb{E}[\cdot]$ denotes the expected value operator.
$\rho(x)$ is the radius of a local \emph{confidence ball} that surrounds the
point $x$: the more uncertain the true location of the estimated filament,
the larger the value of $\rho(x)$. We estimate $\rho(x)$, which is defined 
as a function of the unknown density field $p$ and the unknown filament set 
$R$, by utilizing bootstrap resampling.

\begin{figure*}
\centering
\subfigure[Local uncertainty.]
	{
	\includegraphics[width=2.5 in, height= 2.5 in]{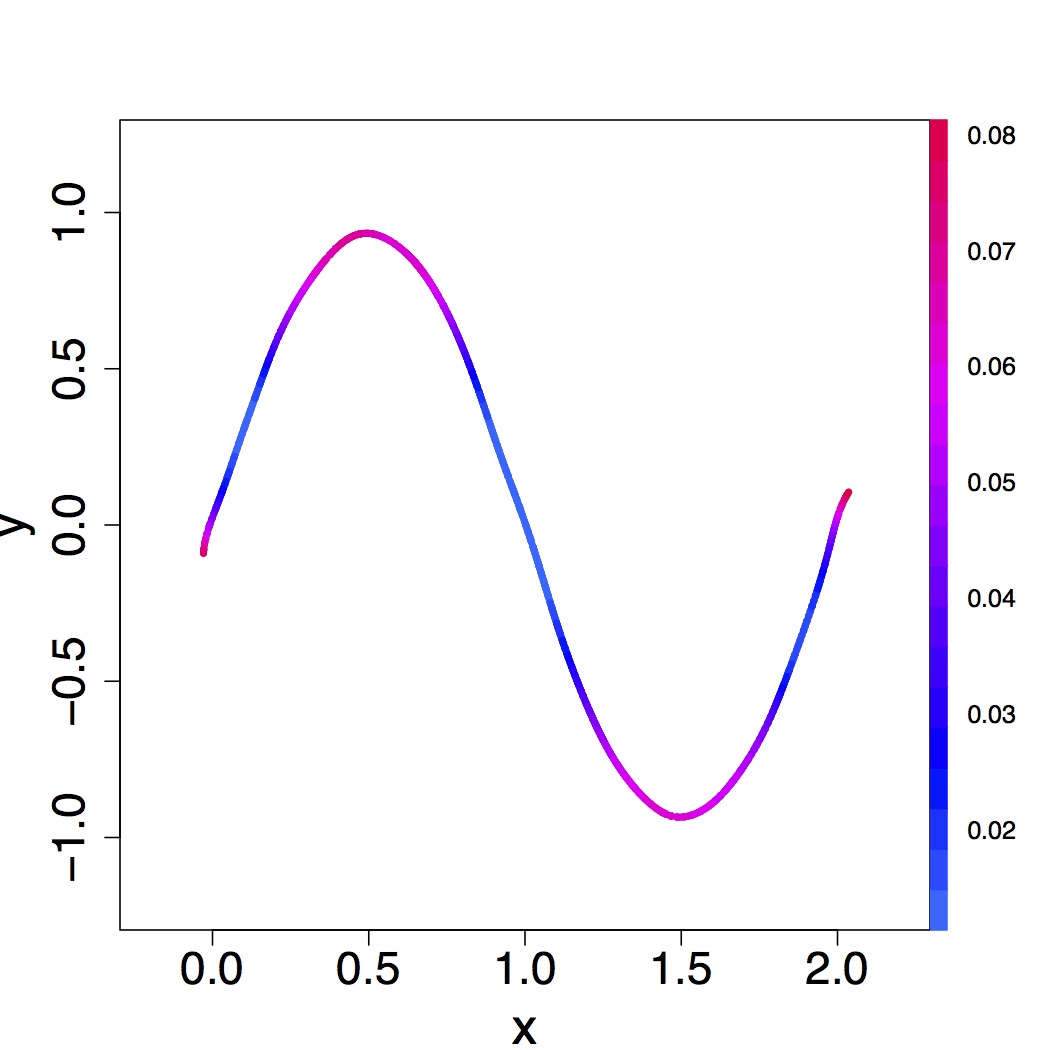}
	}
\subfigure[Uncertainty band.]
	{
	\includegraphics[width=2.5 in, height= 2.5 in]{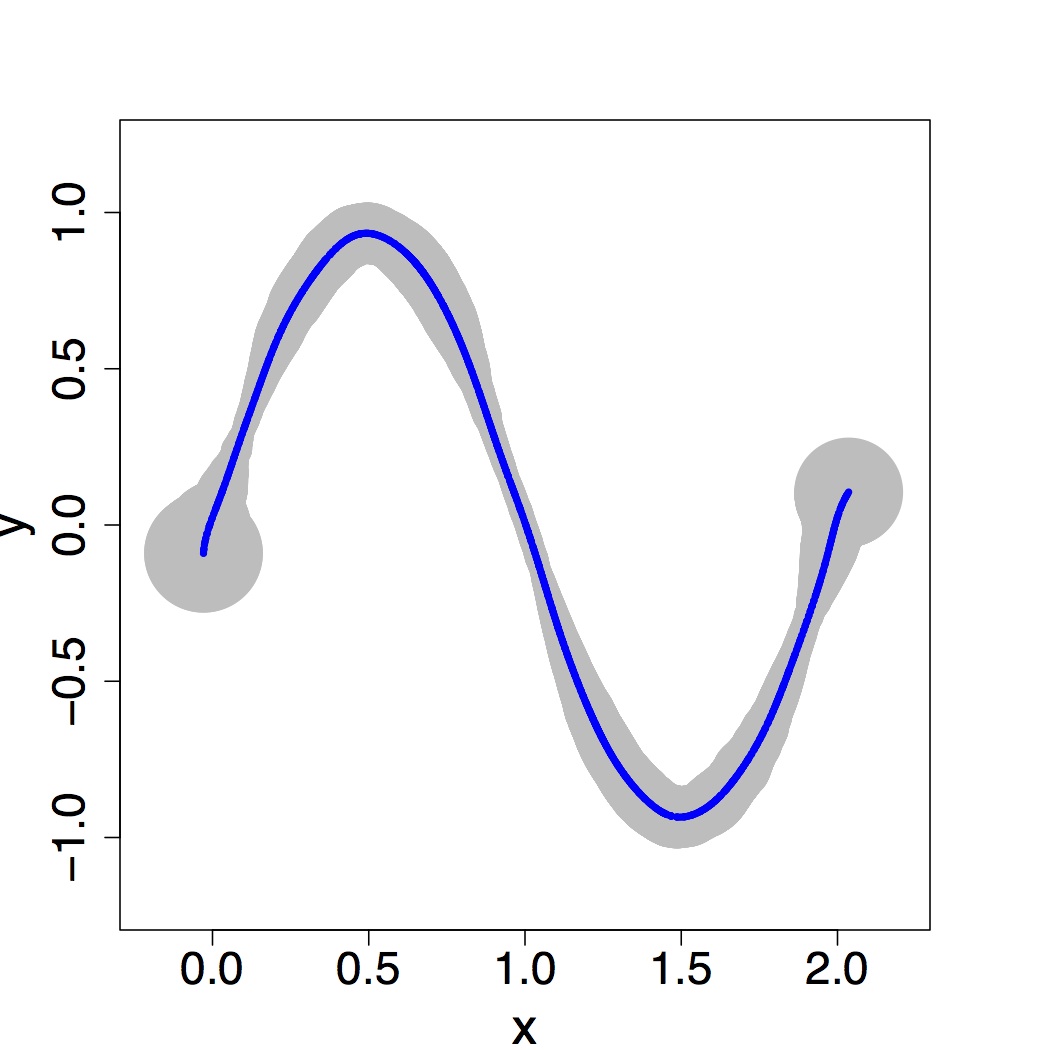}
	}
\caption{An illustration of the uncertainty measure for SCMS. 
In {\bf (a)}, we display the uncertainty measures with different color (red: highly uncertain). 
The unit to the color is the same as $x$ and $Y$ axis.
In {\bf (b)}, we show the uncertainty measures by a gray region around the filament (blue). 
Note that this shows that the SCMS has more uncertainty measures
around the highly curved regions and the end points.}
\label{fig:UMex}
\end{figure*}

In this paper, we consider both the original version of bootstrap 
\citep{Efron1979} and the smooth bootstrap.
The \emph{smooth bootstrap} (see e.g.~\citealt{SilvermanSmooth})
is a variant of the bootstrap that is
useful in functional estimation problems
in which
the bootstrap sample is drawn from the estimated density $\hat p$
instead of the original data. 
When the smoothing kernel is a bivariate Gaussian, we generate
the smooth bootstrap sample via the following two steps:
\begin{itemize}[leftmargin=0.35in]
\item[1.] Generate the bootstrap sample.
\item[2.] Add independent and identically distributed Gaussian noise with variance $h^2$.
\end{itemize}
Unlike the bootstrap, the smooth bootstrap takes into account both the
variance and the bias of filament estimation, but with less precision in
variance estimation with respect to the bootstrap.

Assume we generate $B$ bootstrap samples,
and each of them is denoted as $\{X_1^{*(b)},\cdots,X^{*(b)}_n\}$, $b=1,\cdots, B$.
For each bootstrap sample, say $X_1^{*(b)},\cdots,X^{*(b)}_n$,
we compute the density estimate $\hat{p}^{*(b)}$,
the ridge estimate
$\hat{R}^{*(b)} = \mbox{Ridge}(\hat{p}^{*(b)})$,
and the confidence ball radii $\rho_{(b)}(x)$ for all $x\in\hat R$.
We estimate $\rho(x)$ by adding the $B$ radius estimates in quadrature:
\begin{align}
\hat{\rho}(x) = \sqrt{ \frac{1}{B} \sum_{b=1}^B \rho_{(b)}^2(x) } \,,
\end{align}
In Algorithm \ref{alg::UM} we outline the computational steps that
one must follow to derive $\hat{\rho}(x)$.

Note that calculating the uncertainty measure 
is not part to the SCMS algorithm--we can detect filaments
without using the uncertainty measure.
However, this uncertainty measure is a feature that 
SCMS filaments have.
This measure has a geometric interpretation
and can be consistently estimated.
See \cite{2014arXiv1406.5663C} for more involved discussion.
Note that other filament finders do have have such a statistically consistent
error measurement.

\begin{algorithm*}
\caption{Uncertainty Measure for SCMS}
\begin{algorithmic}
\State \textbf{Input:} Data $\{ X_1,\cdots,X_n\}$. Smoothing bandwidth $h$. Threshold $\tau$.
\begin{itemize}[leftmargin=1.5cm]
\item[\Step 1.] Perform SCMS on $\{ X_1,\cdots,X_n\}$ to detect filaments; 
denote the estimated filaments by $\hat{R}$.
\item[\Step 2.] Generate $B$ bootstrap samples: $X^{*(b)}_1,\cdots,X^{*(b)}_n$ for $b = 1,\cdots,B$.
\item[\Step 3.] For each bootstrap sample, apply SCMS which yields $\hat{R}^{*(b)}$ for $b = 1,\cdots,B$.
\item[\Step 4.] For each $x\in \hat{R}$, calculate $\rho^2_{(b)}(x) = d^2(x,\hat{R}^{*(b)})$, $b=1,\cdots,B$.
\item[\Step 5.] Compute $\hat{\rho}(x) = \left[ \mbox{mean}\{\rho_1^2(x),\cdots,\rho_B^2(x)\} \right]^{1/2}$.
\end{itemize}
\smallskip
\State \textbf{Output:} $\hat{\rho}(x)$.
\end{algorithmic}
\label{alg::UM}
\end{algorithm*}

\subsection{SCMS: Boundary Bias}

When computed with a kernel density estimator as in equation \eqref{eq::KDE}, SCMS filament estimates 
suffer from \emph{boundary bias} within $\sim$ two bandwidths of the edge of the observation
region. This is a systematic deviation from the true filament caused by the density estimator
averaging over a region where no data can be observed, and it can degrade the confidence
band coverage probabilities near the boundary.  One remedy for boundary bias is
to include additional data immediately outside
of the region of interest. Including galaxies within $2h$ of the boundaries
eliminates most of the boundary bias, since very little of the volume
under a bivariate Gaussian kernel lies beyond that point.
If one cannot include additional data points outside the boundaries
(for instance, due to overall survey limits),
then one must be careful when interpreting filaments detected near the
boundaries. 

\subsection{Filament Coverage}
Here we introduce some useful geometric concepts about coverage.
Given two sets $A$ and $B$.
The \emph{coverage} of $B$ by $A$ is defined as
\begin{equation}
\Cov_B(A) = \frac{\mbox{Number of points in } (A\cap B)}{\mbox{Number of points in } B}.
\end{equation}
Note that when $A$ and $B$ are curves, they will contain infinite number of points.
In this case, we will replace `number of points in' by `the length of'.
Similarly, we can define the coverage of $A$ by $B$ as $\Cov_A(B)$.

Given two collections of filaments $R_1$ and $R_2$,
since $R_1$ and $R_2$ are curves so that they may not intersect each other in general
so that the coverage is $0$.
Thus, instead of directly compute their coverage, we consider a flatten version of $R_1$
(and $R_2$ respectively). We define 
\begin{equation}
R_1\oplus r = \{x: d(x,R_1)\leq r\}
\end{equation}
as the r-flatten set of $R_1$.
Then we define the coverage of $R_2$ by $R_1$ as a function of $r$ as
\begin{equation}
\Cov_{R_2}(r; R_1) = \frac{\mbox{Number of points in } (R_2\cap (R_1\oplus r))}{\mbox{Number of points in } R_2}.
\end{equation}
Similarly, we can define $\Cov_{R_1}(r; R_2)$.
The two functions $\Cov_{R_1}(r; R_2)$ and $\Cov_{R_2}(r; R_1)$ 
contain information about the similarity between $R_1$ and $R_2$.

In simulation, we are able to define true filaments, say $R_{true}$,
and we will have an estimate filament, denoted as $\hat{R}_n$.
Then we call the quantity $\Cov_{R_{true}}(r; \hat{R}_n)$ the \emph{true positive coverage}
(ratio of true filaments being covered by estimated filaments)
and we call $1-\Cov_{\hat{R}_n}(r; R_{true})$ the \emph{false positive coverage}
($\Cov_{\hat{R}_n}(r; R_{true})$ is the ratio of estimated filament being covered by 
truth so that 1 minus this ratio is the ratio of false positive). 
See Figure~\ref{fig::TPFP} for an example of true positive and false positive coverage.

Combining the uncertainty measures and the coverage, we
can study the properties of the \emph{uncertainty band.}.
An uncertainty band for a detected filament is simply the union of
the confidence balls computed for each point on the filament, i.e.
\begin{equation}
\hat{U}(k) = \hat{R}\oplus k \hat{\rho}\equiv \bigcup_{x\in\hat{R}}B(x, k\hat{\rho}(x)) \,,
\end{equation}
where $B(x,r) = \{y: \norm{x-y}\leq r\}$ represents the set of points within
a ball centered at $x$ and with radius $r$.
Denote the region within the uncertainty band as $A$.
The coverage for $A$ is then
\begin{equation}
\begin{aligned}
\FCov(A) &= \Cov_{R_{true}}(A)\\ 
&= \frac{\mbox{Number of points in } (A\cap R_{true})}{\mbox{Number of points in } R_{true}}.
\end{aligned}
\end{equation}
One can think of $\FCov(A)$ as the true positive coverage 
using a set $A$.
For instance,
if $\FCov(A)=0.8$, then on average, 80\% of the points on any given true 
filament lie within its associated uncertainty band, and 20\% lie outside
the band. This interpretation of coverage differs from the standard 
interpretation of confidence band coverage,
thus motivating our use of the term ``uncertainty band" instead of 
``confidence band."
Figure \ref{fig::FCov} gives examples of the coverage for
uncertainty bands $\hat{U}(k)$ with $k\in(0{,}3)$ and $n = 250$ and 2500. 
As we observe in Figure \ref{fig::FCov}, the 
coverage percentage depends sensitively on the sample size $n$; thus, we
cannot provide simple rules for converting $k\sigma$ uncertainty bands to 
coverage percentages.


\section{Subspace Constrained Mean Shift: Applications}

\subsection{Voronoi Dataset}	\label{sec::voronoi}
To show the effectiveness of capturing filaments,
we compare the SCMS filaments (density ridges)
to the filaments in the Voronoi model.
The Voronoi model \citep{1994A&A...283..361V}
applies Voronoi tessellation to compute a density estimate 
for galaxies as well as the curvature of that estimate.
Given a curvature estimate, the Voronoi method assigns a class label
to each galaxy, indicating the type of large-scale structure 
to which to associate the galaxy.
There are four possible classes: {\tt cluster}, {\tt filament}, 
{\tt wall}, and {\tt void}.

\begin{figure*}
\centering
\subfigure[All galaxies]
	{
	\includegraphics[width=2 in]{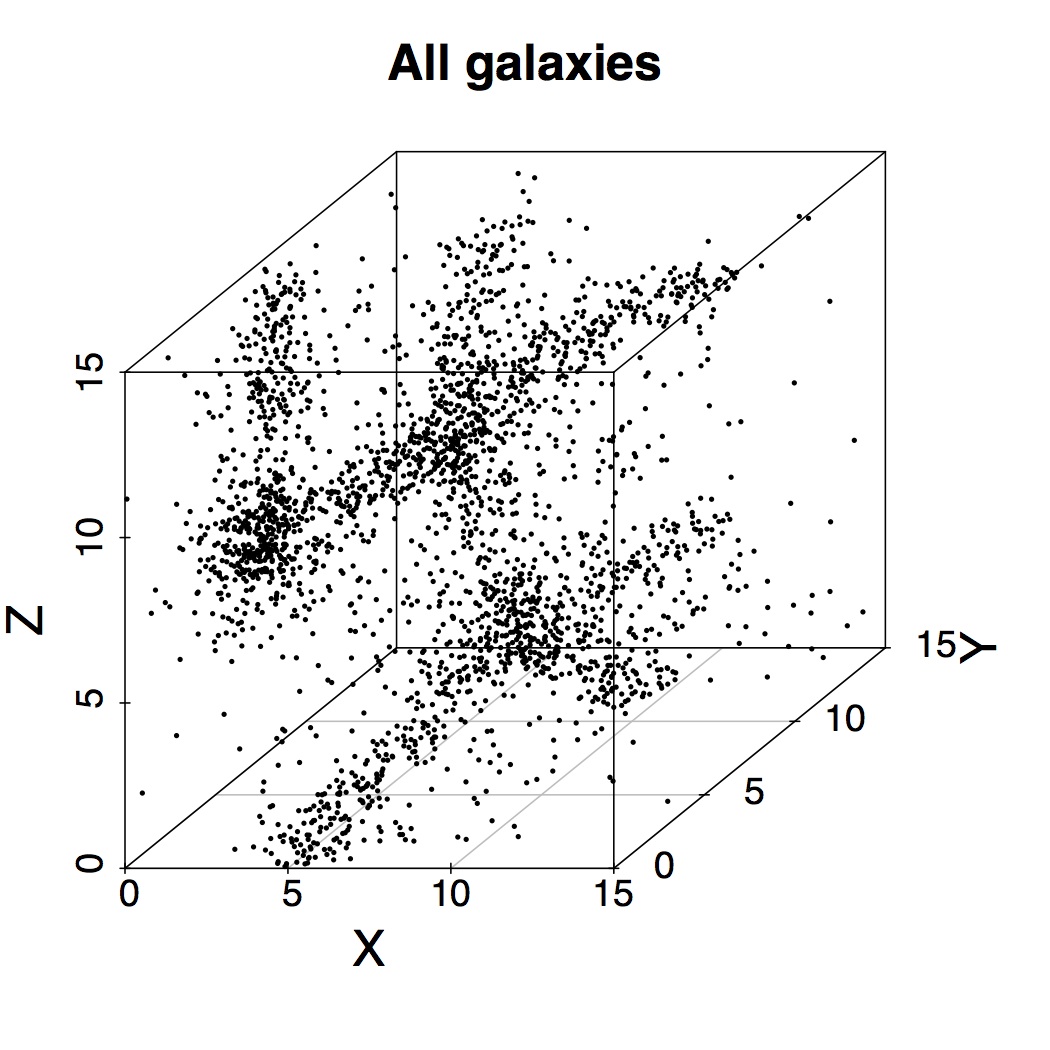}
	}
\subfigure[Galaxies with label $=$ clusters]
	{
	\includegraphics[width=2 in]{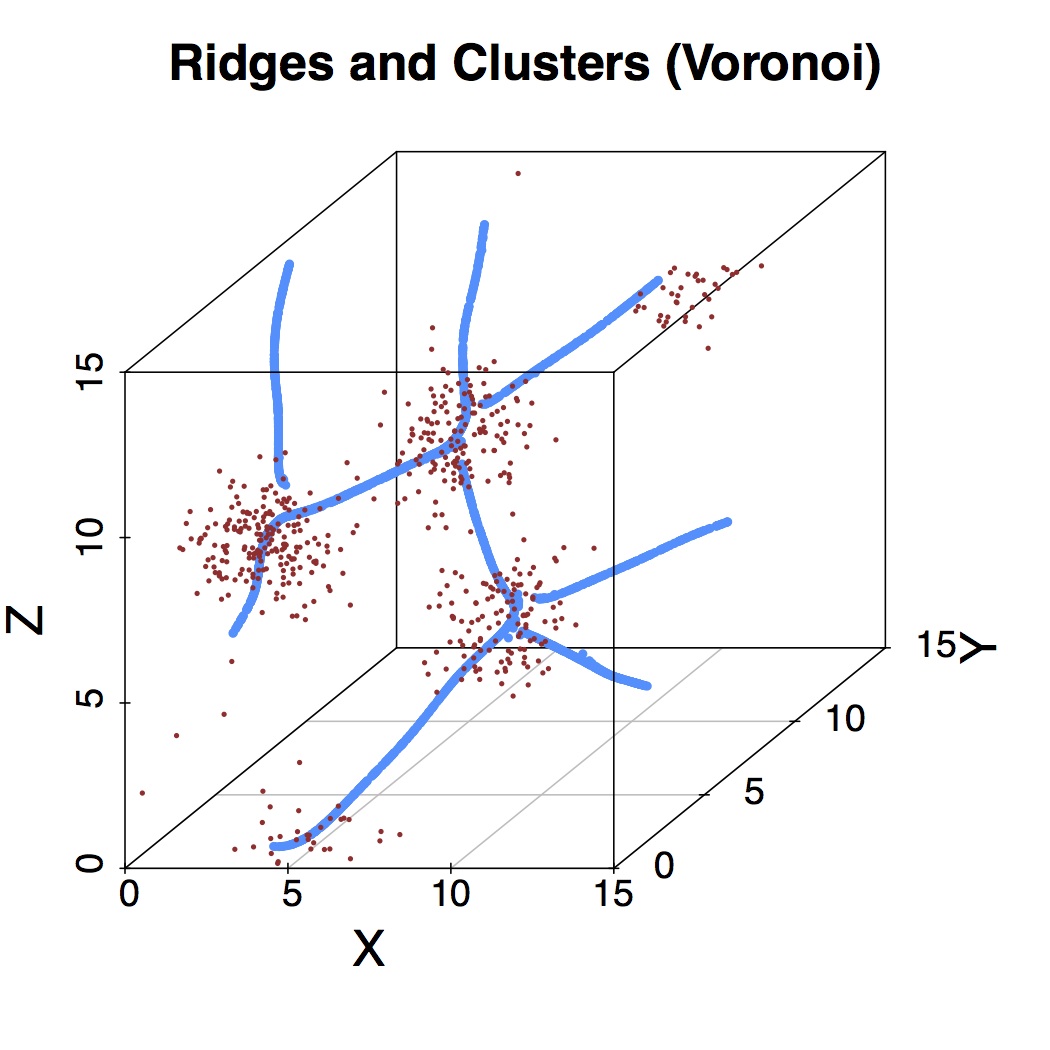}
	}
\subfigure[Galaxies with label $=$ filaments]
	{
	\includegraphics[width=2 in]{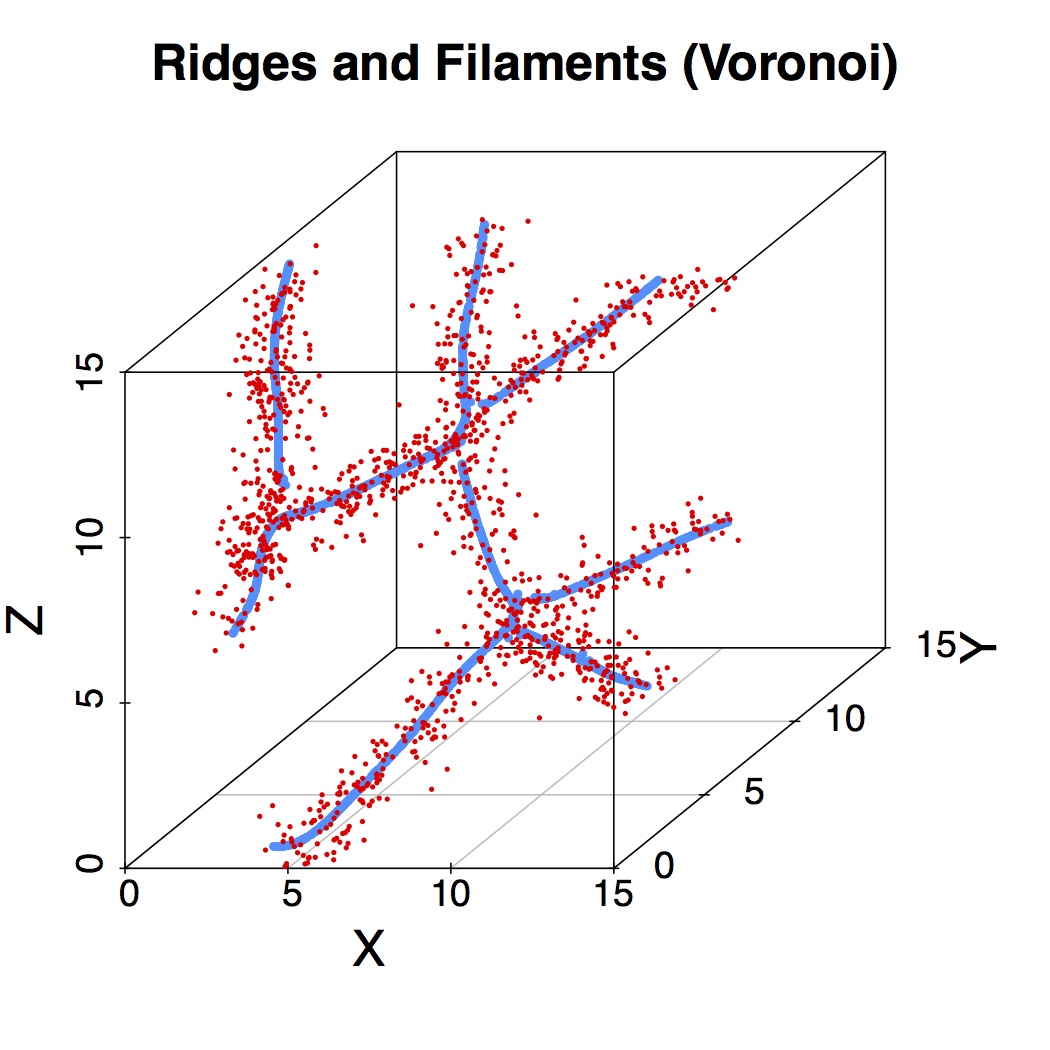}
	}
\subfigure[Galaxies with label $=$ walls]
	{
	\includegraphics[width=2 in]{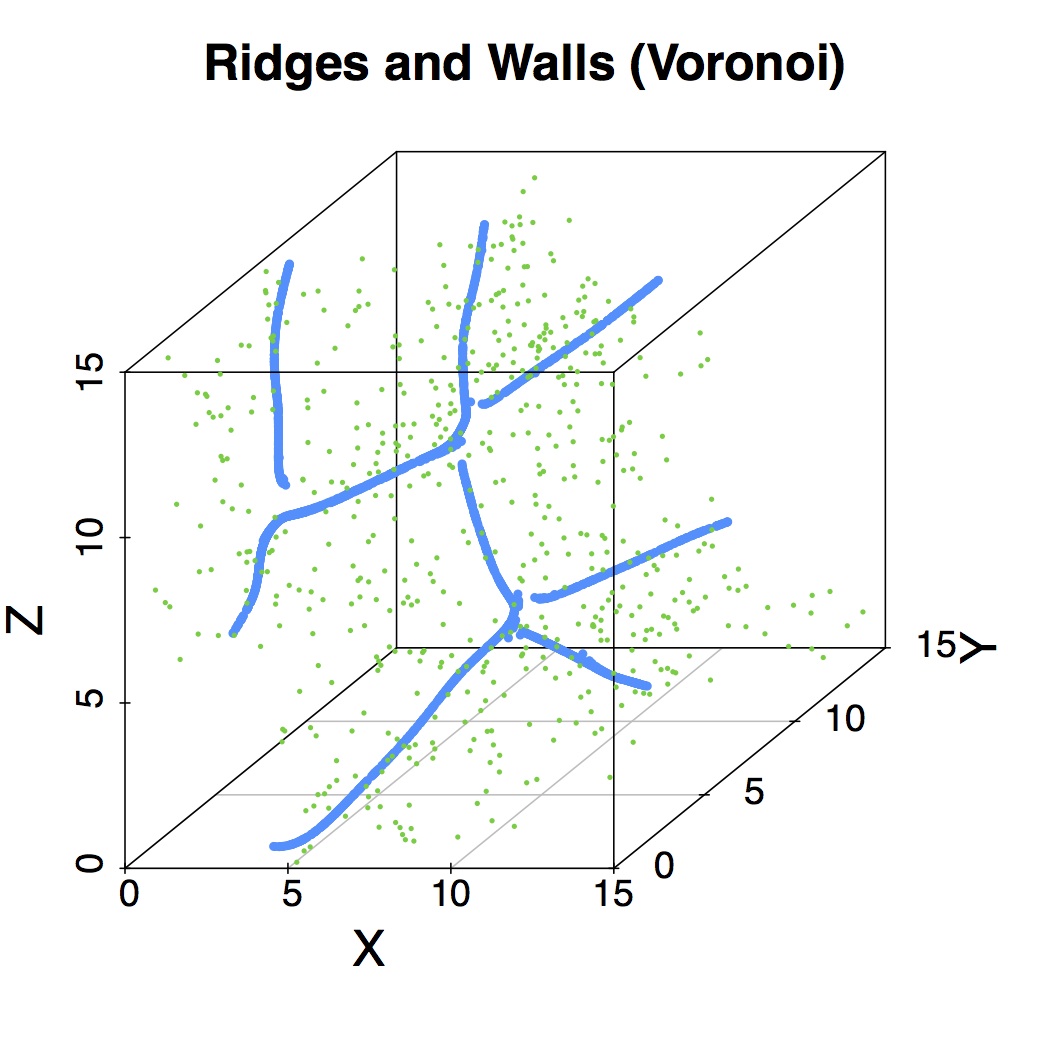}
	}
\subfigure[Galaxies with label $=$ voids]
	{
	\includegraphics[width=2 in]{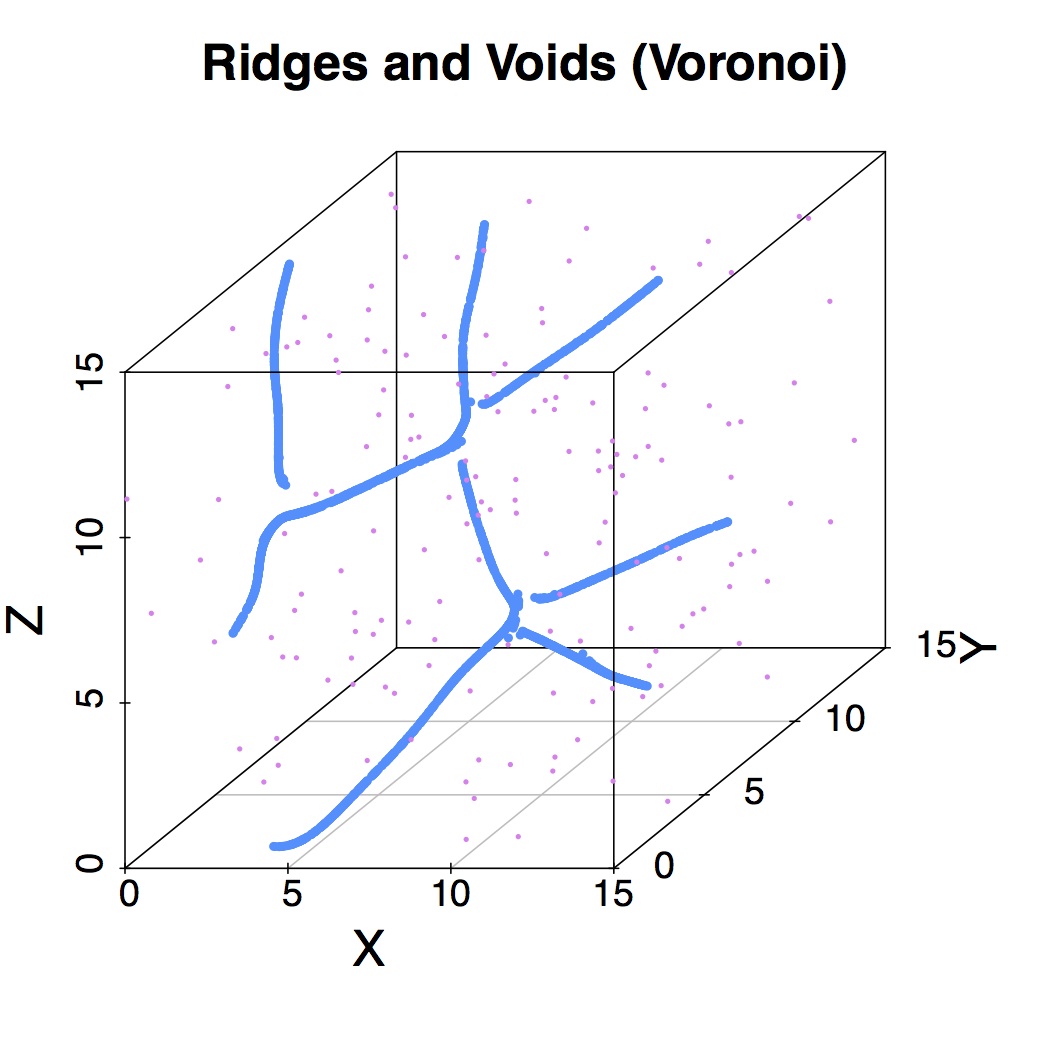}
	}
\caption{
A comparison between density ridges and Voronoi model.
In each panel, the blue curves are density ridges using
all galaxies.
Panel (b)-(e) display the comparison of density ridges
to the Voronoi clusters, filaments, walls and voids.
In panel (c), we see a remarkable similarity between
density ridges and the Voronoi filaments.
}
\label{fig::voro}
\end{figure*}

We use the SCMS algorithm to analyze a simulated dataset 
($256^3$ galaxies, each with a class label, that span 
a $100\times 100\times 100$ Mpc$^3$ box)
generated with the Voronoi model
(M.~A.~Arag\'on-Calvo, private communication).
Figure~\ref{fig::voro} shows a comparison between our
density ridges (blue curves) and galaxies with different class labels
(brown: {\tt cluster}; red: {\tt filament}; green: {\tt wall}; 
pink: {\tt void}). The two methods generate remarkably similar results:
Voronoi clusters (i.e.~galaxies labeled {\tt cluster}) occur at the 
intersection points of density ridges;
Voronoi filaments surround the density ridges; and
Voronoi walls span surfaces on which the density ridges lay.

\begin{figure*}
\centering
	\includegraphics[width=2 in]{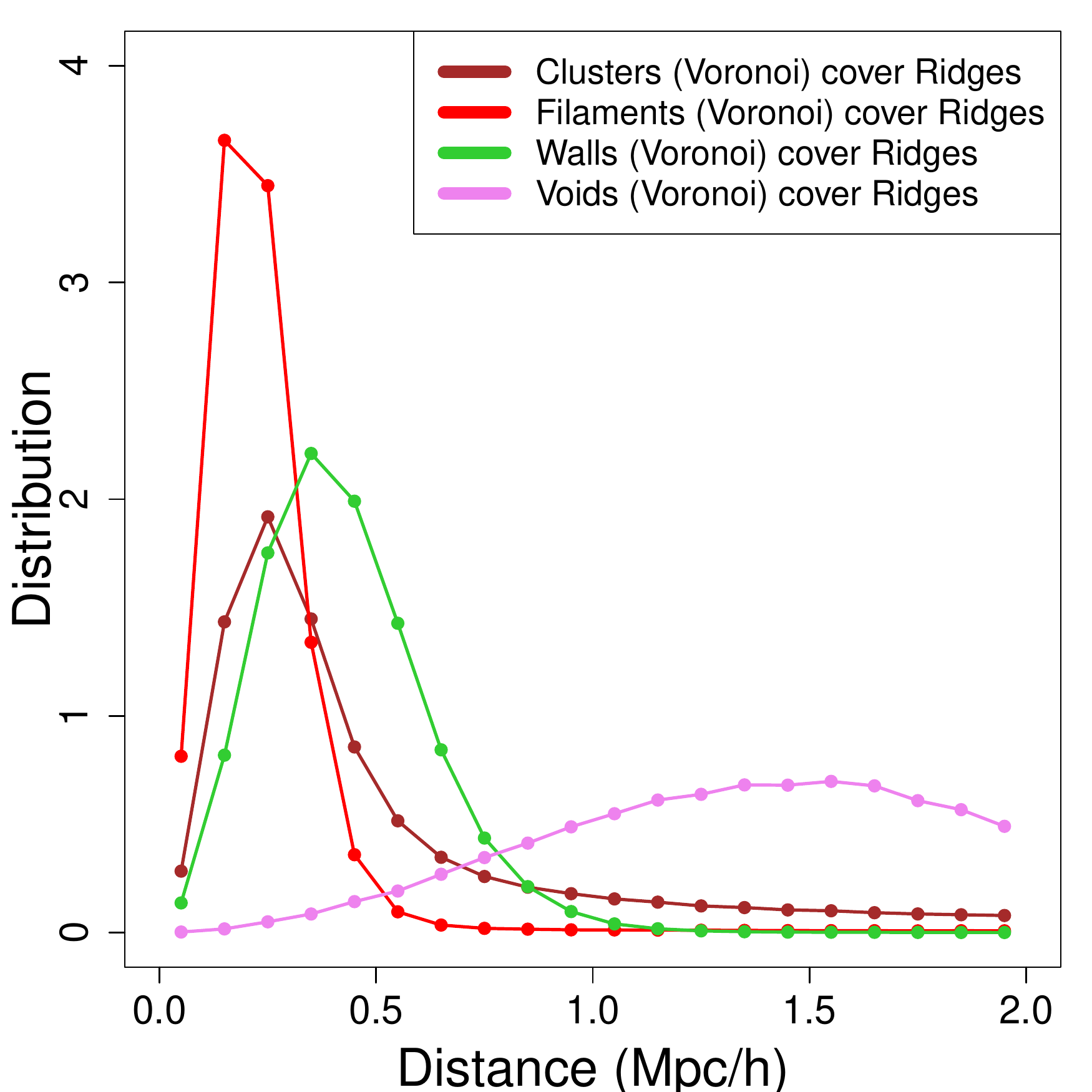}
	\includegraphics[width=2 in]{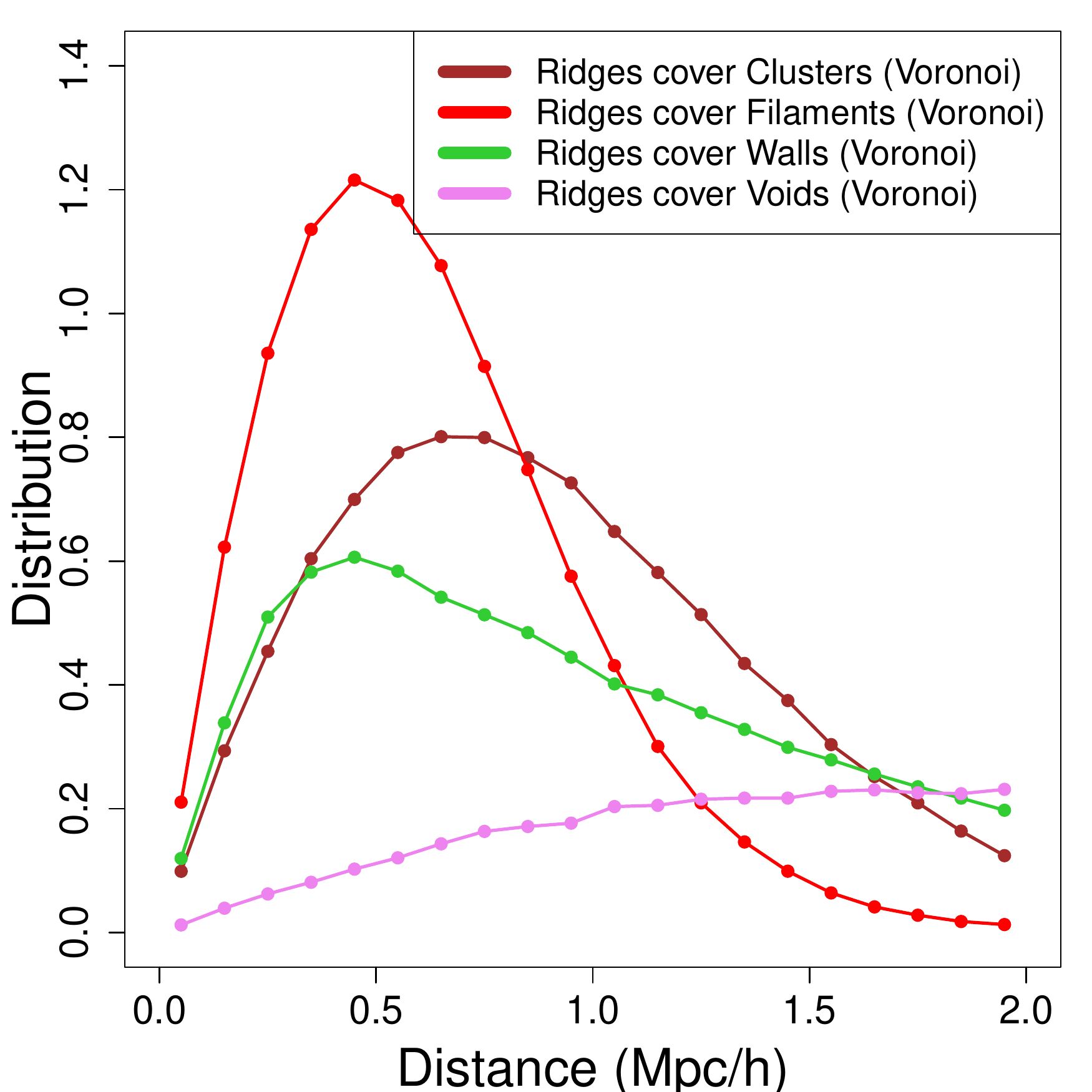}
\caption{
The distributions for projection distances from 
Voronoi-model-derived structures onto density ridges (left panel)
and vice-versa (right panel). Both panels indicate that density ridges trace
structures most similar to Voronoi filaments.
}
\label{fig::voro2}
\end{figure*}

To further quantify the association between
density ridges and each Voronoi model class,
we study their projection distances onto each other.
Note that the distribution of projection distances is related
to filament coverage; further discussion of this
may be found in \cite{2015arXiv150602278C}.
Figure~\ref{fig::voro2} displays the distributions of projection distances.
In both panels, we see that the distribution for ridges versus the 
Voronoi filaments peaks at distances $\lesssim$ 1 $h^{-1}$Mpc
This indicates that the density ridges and the Voronoi filaments
are very similar. On the other hand, the projection distances from
the density ridges increases as we consider clusters, walls, and voids;
the distributions exhibit increasing positive skewness.

\subsection{P3M N-body Simulation}
\label{sec::simulation}

To further demonstrate the efficacy of SCMS, 
we apply it to
P3M N-body simulations from \cite{2015arXiv150702685T},
which assume a $\Lambda$CDM cosmology with
$\Omega_m=0.3$, $\Omega_l=0.7$, $\Omega_b=0.045$, $h=0.7$, $\sigma_8=0.8$ 
and $n_s=0.96$. 
Each side of the simulation box is of length 1 Gpc/h, 
and each contains $2048^3$ particles. 

\begin{figure*}
\centering
\subfigure[Density field and true filaments]
	{
	\includegraphics[width=2.5 in, height= 2.5 in]{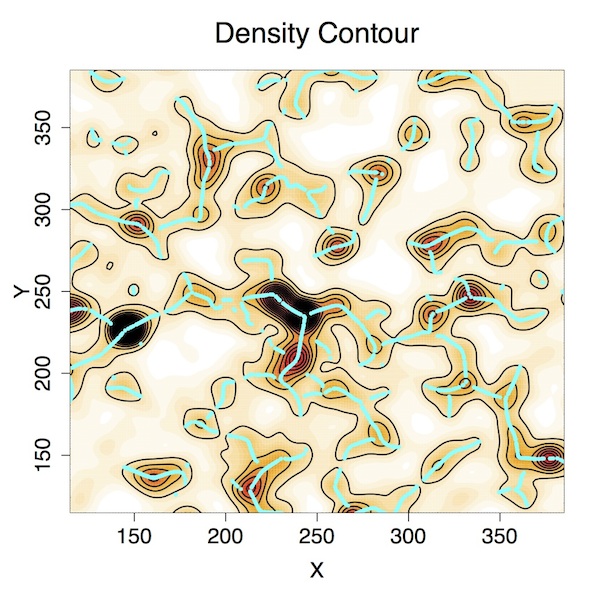}
	}
\subfigure[n=250 (CMASS)]
	{
	\includegraphics[width=2.5 in, height= 2.5 in]{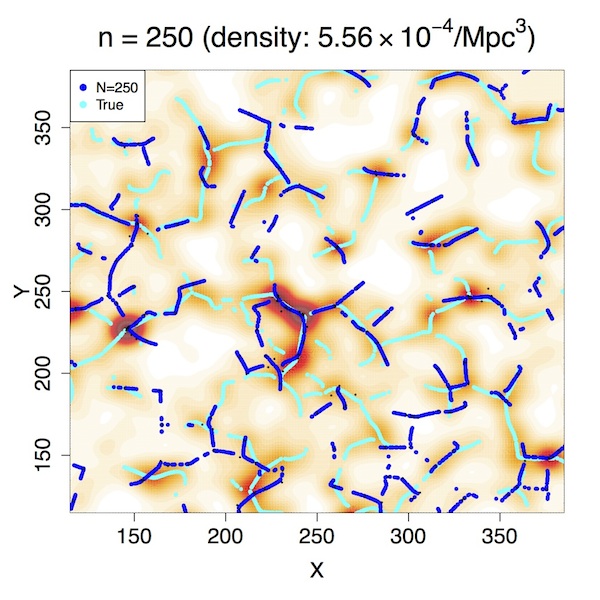}
	}
\subfigure[n=2500 ({\it WFIRST})]
	{
	\includegraphics[width=2.5 in, height= 2.5 in]{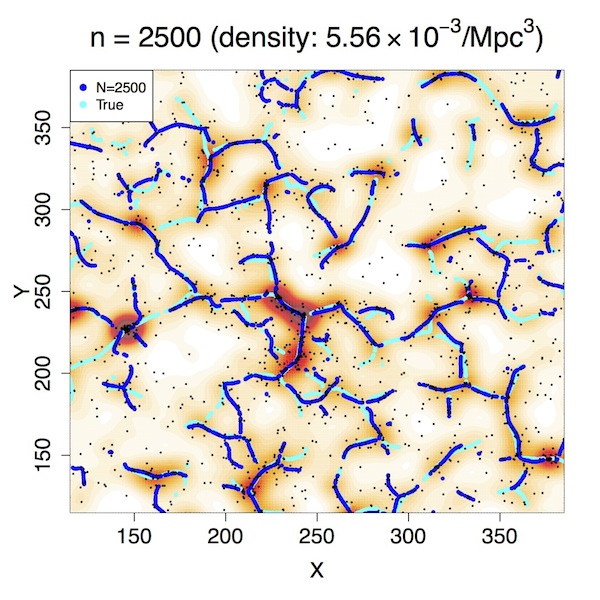}
	}
\subfigure[n=10000]
	{
	\includegraphics[width=2.5 in, height= 2.5 in]{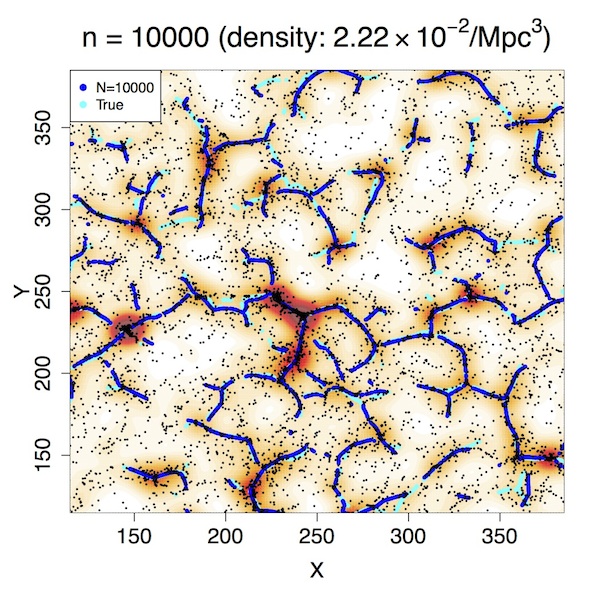}
	}
\caption{A simulated example to show the consistency of SCMS.
This data is a slice of an N-body simulation in a box; the unit of $X$ and $Y$ axes is Mpc/h.
We take a slice with width $5$ Mpc/h.
The original sample contains $88{,}406$ particles. 
The color contour is the galaxy density field from the original sample with smoothing
parameter $h=5$ and
the true filaments (cyan)
are density ridges of this density field.
We subsample under various sizes. The blue curves are estimated filaments
based on the subsample (black dots). One can see a clear pattern; as sample size 
(for the subsample) increases, the estimated filaments are closer to the true filaments.
See Section \ref{sec::simulation} for more details.
}
\label{fig::simN1}
\end{figure*}

\begin{figure*}
\centering
\subfigure[True positive coverage.]
	{
	\includegraphics[width=2.5 in, height= 2.5 in]{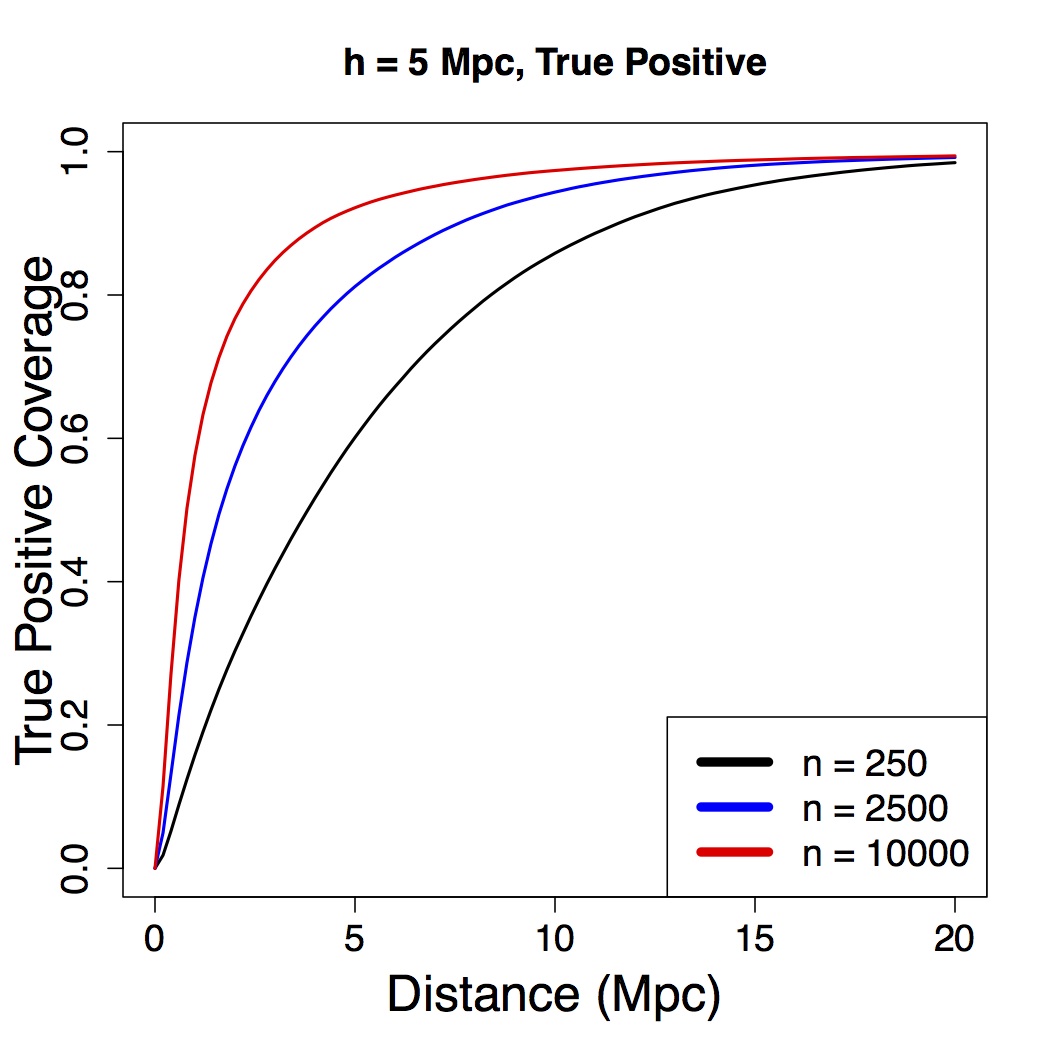}
	}
\subfigure[False positive coverage]
	{
	\includegraphics[width=2.5 in, height= 2.5 in]{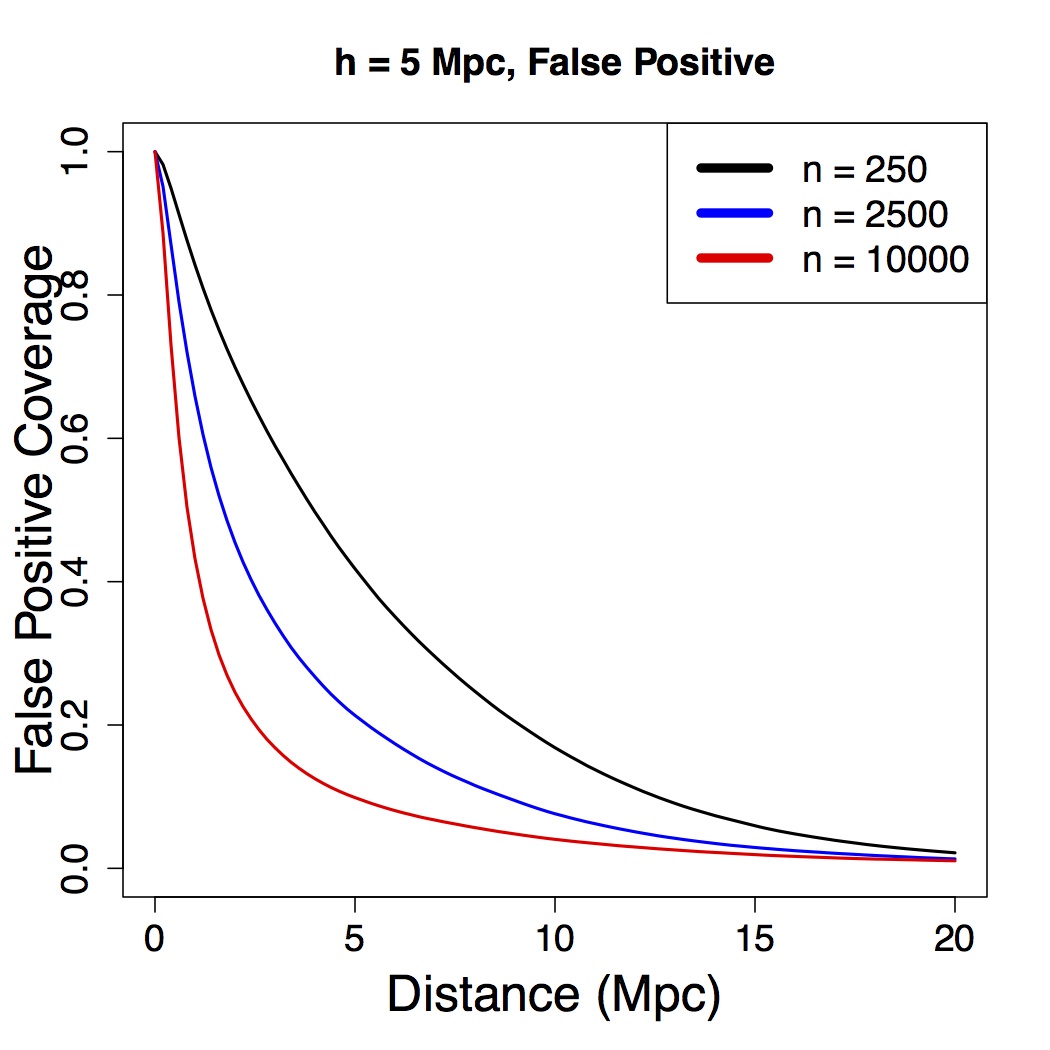}
	}
\caption{True positive and false positive coverage. 
As can be seen, for all distances, the true positive coverage increases
with sample size, whereas the false positive coverage decreases.
}
\label{fig::TPFP}
\end{figure*}

In Figure~\ref{fig::simN1}, we demonstrate
that, as sample size increases,
SCMS outputs filament estimates that 
are closer to the true filaments (defined by the true density function);
the uncertainty measures capture SCMS errors
due to the sampling variability.
We take a slice of the full simulation data 
($x,y\in[125{,}375]$ Mpc/h and $z\in[100{,}105]$ Mpc/h)
and smooth the data with smoothing bandwidth $h=5$ (recommended by 
the selection rule in Appendix \ref{sec::parameters} with $A_0=0.4$)
to get the density function
and the filaments (cyan curves). 
Figure~\ref{fig::simN1}(a) shows a contour plot for the density function.
The original sliced data contains $88{,}406$ points (gray dots).
We downsample to get three different subsamples;
each contains $250/2500/10000$ particles.
For each subsample (black dots), we apply SCMS
to detect filaments (blue curves).
Note that the convergence phenomena of Figure~\ref{fig::simN1} are further quantified by
the true positive and false positive coverage plot in Figure~\ref{fig::TPFP}.

Note that the sparsest subsample $n=250$ has a galaxy number density
$5.56\times 10^{-4}$ Mpc$^{-3}$
which is similar to
the number density observed in SDSS CMASS data ($\sim4 \times 10^{-4}$
Mpc$^{-3}$).
The future survey
\emph{Wide-Field Infrared Survey Telescope (WFIRST)},\footnote{
http://wfirst.gsfc.nasa.gov/}
a NASA mission with science objectives in exoplanet exploration, dark energy 
research and galactic and extragalactic surveys,
will observe a number density similar to the $n=2500$ subsample ($\sim5.56\times 10^{-3}$ Mpc$^{-3}$).

\begin{figure*}
\centering
\subfigure[Filament coverage.]
	{
	\includegraphics[width=2.5 in, height= 2.5 in]{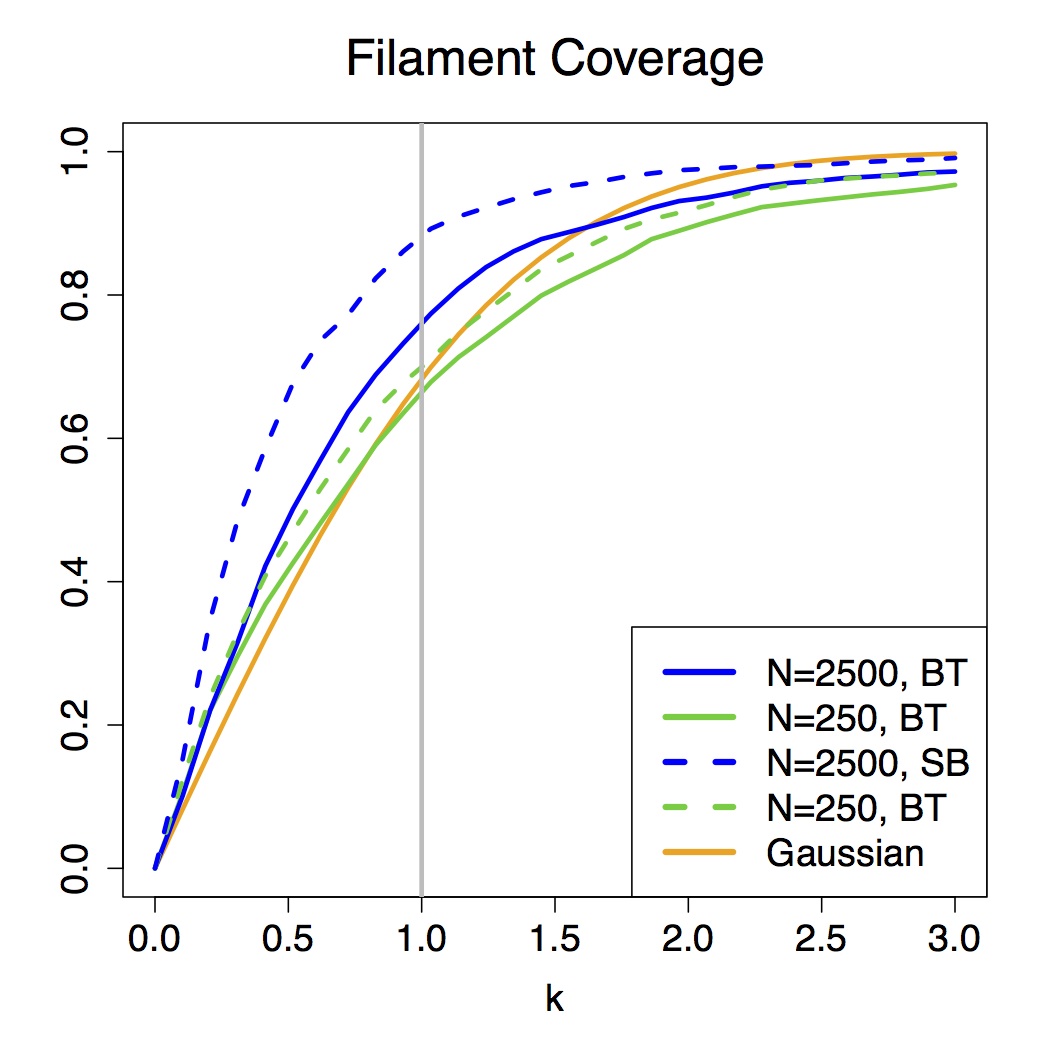}
	}
\subfigure[$N=2500$, smooth bootstrap]
	{
	\includegraphics[width=2.5 in, height= 2.5 in]{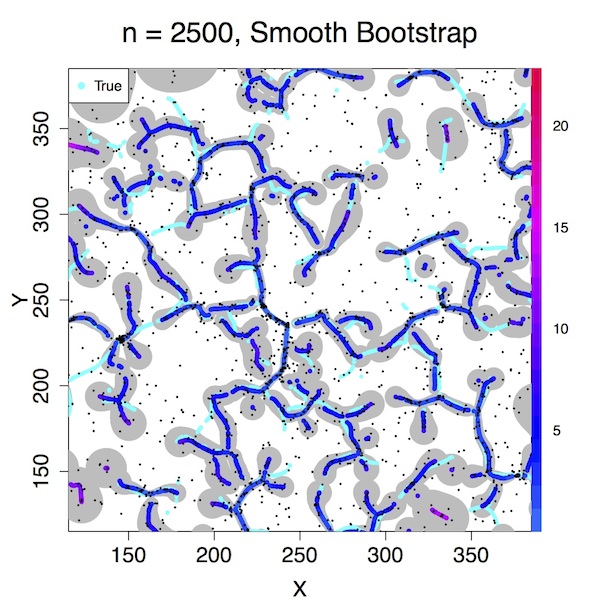}
	}
\caption{Filament coverage based on the uncertainty measure. 
{\bf (a)} The filament coverage $\FCov(\hat{U}(k))$ as a function of $k$ (x-axis). 
We also provide the coverage for Gaussian distribution (probability being within $k\sigma$
to the center of Gaussian) as a reference.
{\bf (b)} Visualizing the uncertainty by color and a confidence set for the 
subsample with $n=2{,}500$ with the uncertainty measure
estimated via the smooth bootstrap.
The cyan curves are the true filaments.
Note that the gray regions are $\hat{U}(k), k=1$,
equivalent to the error bar for $1\times\sigma$, based on 
the smooth bootstrap estimate.
From panel (a), we know that the gray regions contain about $85\%$
true filaments (cyan curves).
The unit to the color in uncertainty band is Mpc, the same as $X$ and $Y$ axes.}
\label{fig::FCov}
\end{figure*}

We show the uncertainty measures and filament coverage
for $n=2500$ in Figure~\ref{fig::FCov}.
We plot filament coverage for confidence regions $\hat{U}(k)$ 
for $k\in(0{,}3)$ in Figure~\ref{fig::FCov}(a), where $n=250$ and $2500$, and
where $\hat{\rho}$ is estimated by 
both the bootstrap (BT) and the smooth bootstrap (SB). This range
contains sample sizes that are in line with both CMASS ($n \approx$ 250) and 
{\it WFIRST} ($n \approx$ 2500) data.
We observe that filament coverage is, as noted above, sensitive to the
sample size $n$ and that the smooth bootstrap provides considerably 
more conservative confidence bands, particularly for $k \lesssim 2$.
The gray regions displayed in Figure~\ref{fig::FCov}(b)
are the smooth bootstrap confidence regions 
$\hat{U}(1)$, which we estimate contain 
85\% of the true filaments (cyan curves).

\begin{figure*}
\centering
\subfigure[]
	{
	\includegraphics[width=2.5 in, height= 2.5 in]{figures/D2_001}
	}
\subfigure[]
	{
	\includegraphics[width=2.5 in, height= 2.5 in]{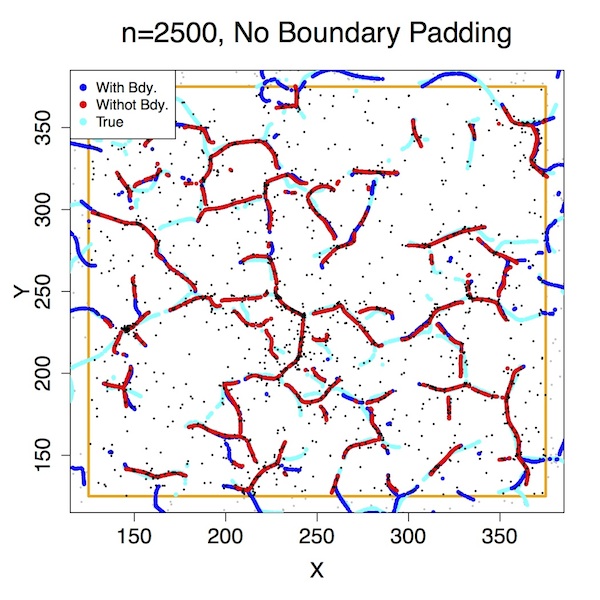}
	}
\subfigure[Uncertainty Measure with Boundary Points]
	{
	\includegraphics[width=2.5 in, height= 2.5 in]{figures/D2_002}
	}
\subfigure[Uncertainty Measure without Boundary Points]
	{
	\includegraphics[width=2.5 in, height= 2.5 in]{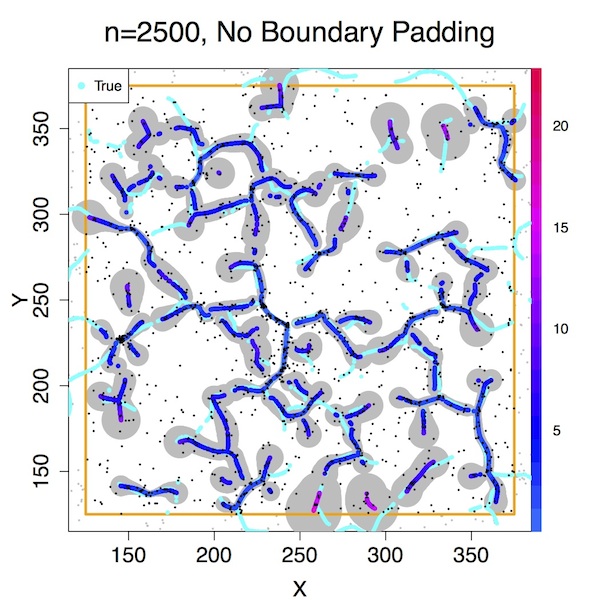}
	}
\caption{Simulated example with sample size $2500$ that demonstrates 
the boundary bias of SCMS.
To demonstrate this bias, 
we remove points outside the orange rectangle 
(the so-called boundary points).
(a) and (b): Comparisons between SCMS with boundary points (blue) 
and SCMS without boundary points (red). 
As can be seen, the bias (between red and blue curves)
is large for filaments whose distance to the boundary 
are less than $10$ Mpc/h ($2\times $ smoothing bandwidth $h$).
(c) and (d): Uncertainty measure for the filaments with and without boundary points.
Notice that in (d), filaments near the boundary tend to have much higher uncertainty.
The unit to the color in uncertainty band is Mpc, the same as $X$ and $Y$ axes.}
\label{fig::bdy}
\end{figure*}

Figure \ref{fig::bdy}
illustrates the effect of boundary bias in the $n=2500$ subsample
by comparing the estimates and uncertainties with padded and unpadded data near the boundary.
Panels (a) and (b) show the boundary bias.
Note that the red curves are filaments estimated 
by using only points within the boundary (given by the orange rectangle).
The blue curves are filaments detected by SCMS with 
boundary points (i.e. points outside the orange rectangle).
As can be seen, the estimation of filaments without boundary
data (red curves) becomes more inaccurate as we approach the boundary.
The boundary bias occurs for filaments with distances less than 10 Mpc/h
($2$ times smoothing parameter $h$) to the boundaries.
The uncertainty measures also show the influence of boundary bias.
Figure \ref{fig::bdy}(c) and \ref{fig::bdy}(d) exhibit the uncertainty measures for filaments estimated
with and without boundary points.
As expected, the uncertainty measures in panel (d) increase as we move close to the boundary.

\subsection{Sloan Digital Sky Survey}
\label{sec::data}

\subsubsection{Data}

We further demonstrate the efficacy of SCMS by 
applying it to data from Data Release 12 \citep{2015arXiv150100963A}
of the Sloan Digital Sky Survey \citep[SDSS;][]{2000AJ....120.1579Y}.
Together, SDSS I, II
\citep{2009ApJS..182..543A}, and III \citep{2011AJ....142...72E} used a drift-scanning
mosaic CCD camera \citep{1998AJ....116.3040G} to image over one third of the sky
($14{,}555$ square degrees) in five photometric bandpasses
\citep{1996AJ....111.1748F,2002AJ....123.2121S,2010AJ....139.1628D} to a limiting magnitude of $r\simeq 22.5$,
using the dedicated 2.5-m Sloan Telescope \citep{2006AJ....131.2332G} located at
Apache Point Observatory in New Mexico. 
The imaging data were
processed through a series of pipelines that perform astrometric
calibration \citep{2003AJ....125.1559P}, photometric reduction \citep{2001ASPC..238..269L}, and
photometric calibration \citep{2008ApJ...674.1217P}. All of the imaging was reprocessed as part
of SDSS Data Release 8 (2011ApJS..193...29A; \citealt{2011ApJS..193...29A}).

The Baryon Oscillation Spectroscopic Survey (BOSS) has obtained spectra and redshifts for 1.35 million
galaxies over a footprint covering $10{,}000$ square
degrees. These galaxies are selected from the SDSS 2011ApJS..193...29A imaging and are
being observed together with $160{,}000$ quasars and approximately
$100{,}000$ ancillary targets.  The targets are assigned to tiles of
diameter $3^\circ$ using a 2003AJ....125.2276B algorithm that is adaptive to the
density of targets on the sky \citep{2003AJ....125.2276B}. Spectra are obtained
using the double-armed BOSS spectrographs \citep{2013AJ....146...32S}. 
Each observation is performed in a series of $900$-second exposures,
integrating until a minimum signal-to-noise ratio is achieved for the
faint galaxy targets. This ensures a homogeneous data set with a high
redshift completeness of more than $97$ percent over the full survey
footprint. Redshifts are extracted from the spectra using the methods
described in \cite{2012AJ....144..144B}. A summary of the survey design appears
in \citet{2011AJ....142...72E}, and a full description is provided in \citet{2013AJ....145...10D}.

BOSS selects two classes of galaxies to be targeted for spectroscopy using SDSS 2011ApJS..193...29A imaging: `LOWZ' and `CMASS' (we refer the reader to \citet{And13} for further description of these classes). 
For the LOWZ sample, 
the effective redshift is $z_{\rm eff} =0.32$, slightly lower than that of the
SDSS-II luminous red galaxies (LRGs)
as we place a redshift cut $z<0.43$. 
The CMASS selection yields a sample with a median redshift $z = 0.57$ and a
stellar mass that peaks at $\log_{10}(M/M_{\odot}) = 11.3$ \citep{2013MNRAS.435.2764M}.
Most CMASS targets are central galaxies residing in dark matter haloes of mass
$\sim10^{13}h^{-1}M_\odot$.

We test SCMS using two slices of data: at low and high redshift.
The low-$z$ dataset comprises $1{,}158$ galaxies in the volume
\begin{align*}
135^{\circ}\leq \mbox{RA}\leq 175^{\circ}~~~5^{\circ}\leq \delta \leq 45^{\circ}~~~0.235\leq z \leq 0.240
\end{align*}
while the high-$z$ dataset lies in the volume
\begin{align*}
135^{\circ} \leq \mbox{RA}\leq 175^{\circ}~~~5^{\circ}\leq \delta \leq 45^{\circ}~~~0.530\leq z \leq 0.535
\end{align*}
and contains $4{,}678$ galaxies.
Both samples have a very thin redshift range $\Delta z = 0.005$ 
(the corresponding comoving distance is around $14-21$ Mpc) 
so that their constituent galaxies may be viewed as lying 
on a two-dimensional surface with coordinates (RA,$\delta$).

There are two principal reasons for our choice to perform a 
two-dimensional analysis of the SDSS data.
The first is that there is too large a change in the number density of 
detected galaxies over the SDSS redshift range.
The SCMS algorithm incorporates kernel density estimation to locate
density ridges, and KDE requires a fixed smoothing parameter $h$.
However, in low-density regions, $h$ should be large to obtain reliable 
results, while in high-density regions, $h$ has to be small so as to not
oversmooth the point cloud. The second reason is that when
$z>0.2$, the number density is very low, and performing a three-dimensional
analysis will produce results with large statistical errors due to the 
small sample size. Lower-dimensional analyses result in decreased
statistical error; see e.g.~\citealt{AllNonPar}.

\subsubsection{Results}

\label{sec::filamentmap}

\begin{figure*}
\centering
\subfigure[The original data.]
	{
	\includegraphics[width=2.5 in, height= 2.5 in]{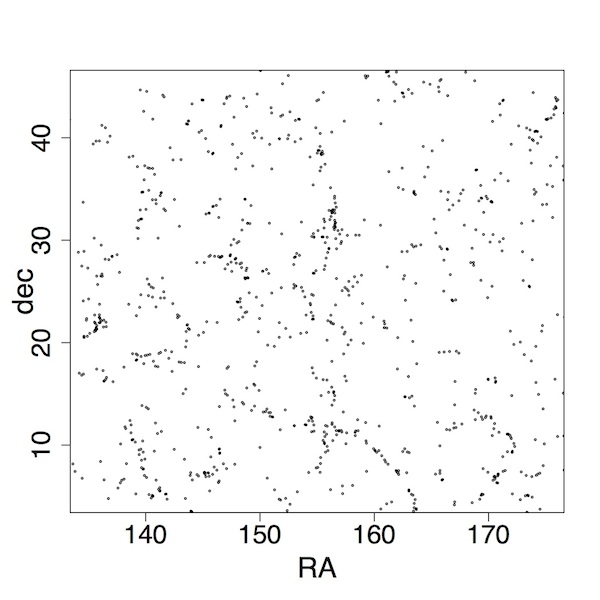}
	}
\subfigure[Filament detection.]
	{
	\includegraphics[width=2.5 in, height= 2.5 in]{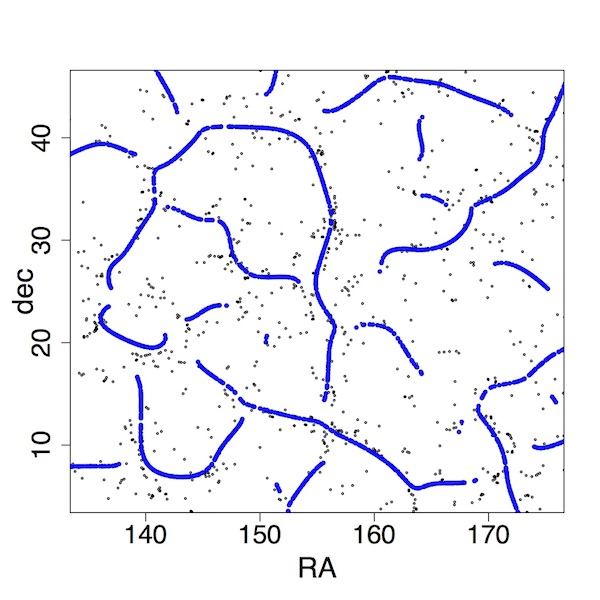}
	}
\caption{Application of SCMS to low-$z$ data ($z=0.235-0.240$). 
The blue curves are filaments detected by SCMS.}
\label{fig::FE1}
\end{figure*}

\begin{figure*}
\centering
\subfigure[The original data.]
	{
	\includegraphics[width=2.5 in, height= 2.5 in]{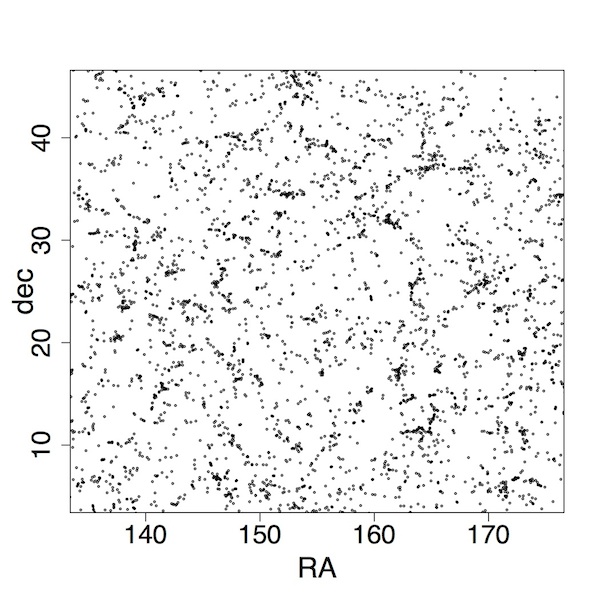}
	}
\subfigure[Filament detection.]
	{
	\includegraphics[width=2.5 in, height= 2.5 in]{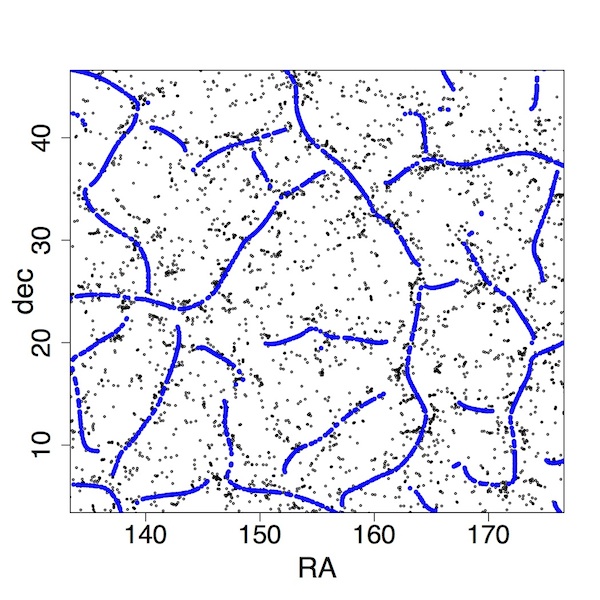}
	}
\caption{Application of SCMS to high-$z$ data ($z=0.530-0.535$). 
The blue curves are filaments detected by SCMS.}
\label{fig::FE2}
\end{figure*}

We apply SCMS to the low-$z$ data with
smoothing bandwidth $h=2.50\degr$ (41.8 Mpc) and threshold 
level $\tau = 1.02\times 10^{-3}$; 
we display our results in Figure~\ref{fig::FE1}.
For the high-$z$ data, $h$ and $\tau$ 
are $2.03\degr$ (71.1 Mpc) and $\tau = 7.52\times10^{-4}$, respectively;
we display our results in Figure~\ref{fig::FE2}.
Note that we have included additional galaxies within $5$ degrees
of our selected window to mitigate boundary bias.

As can be seen in Figures~\ref{fig::FE1} and \ref{fig::FE2}, 
SCMS filament estimates capture high density regions and they exhibit 
one-dimensional, nearly connected structures. 
In addition, SCMS yields smooth filaments; 
most filament estimators do not output such smooth structures
\citep[cf.][]{Stoica2007,2005A&A...434..423S,2011MNRAS.414..350S,Aanjaneya2012,2013arXiv1305.1212L}. 
We note that the filaments detected by SCMS will not actually
connect with each other;
points on merging filaments have eigengap $\beta$ 
(equation \ref{eq::eigengap}) that asymptote toward $0$, 
making the density ridge ill-defined since
the first and second eigenvalues become equal.
We note that in both figures there are possibly spurious filaments;
for instance, in Figure~\ref{fig::FE1}, 
at (RA,$\delta$)$= (165^{\circ}{,}40^{\circ})$ and $(165^{\circ}{,}20^{\circ})$,
we see filaments that are associated with a relatively small number of 
galaxies. As we demonstrate below, these putative filaments have
higher estimates of uncertainty.

\begin{figure*}
\centering
\subfigure[The bootstrap estimate.]
	{
	\includegraphics[width=2.5 in, height= 2.5 in]{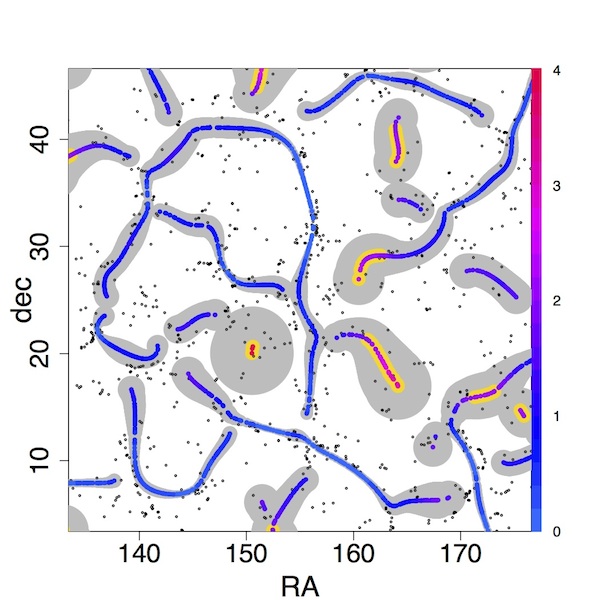}
	}
\subfigure[The smooth bootstrap estimate.]
	{
	\includegraphics[width=2.5 in, height= 2.5 in]{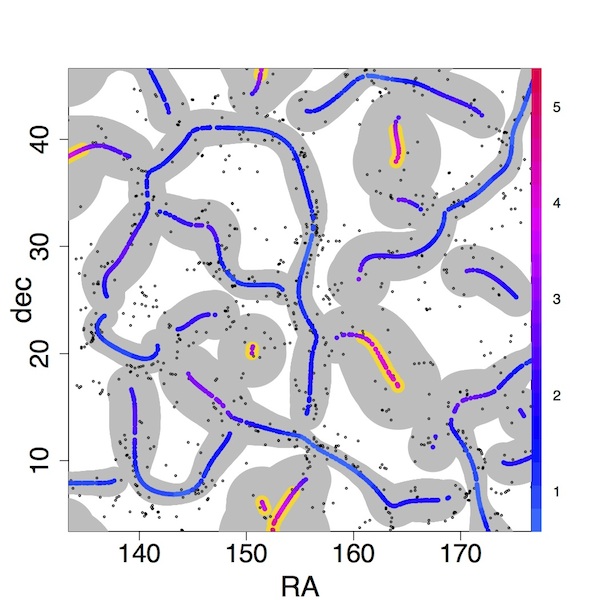}
	}
\caption{Local uncertainty estimates for our low-$z$ SDSS dataset ($z=0.235-0.240$). We display the amount of uncertainty via color (red: high) and a confidence band in (a),(b) using both ordinary bootstrap and the smooth bootstrap. 
The filament points surrounded by yellow colors are those with high uncertainty
and are declared as `unstable'.
Based on the simulation result in Figure~\ref{fig::FCov}, we expect 
that the gray regions in plot (a) contain about $50\%$ true filaments and in 
(b) contain $85\%$ true filaments. 
The unit to the color in uncertainty band is degree.}
\label{fig::LU1}
\end{figure*}

\begin{figure*}
\centering
\subfigure[The bootstrap estimate.]
	{
	\includegraphics[width=2.5 in, height= 2.5 in]{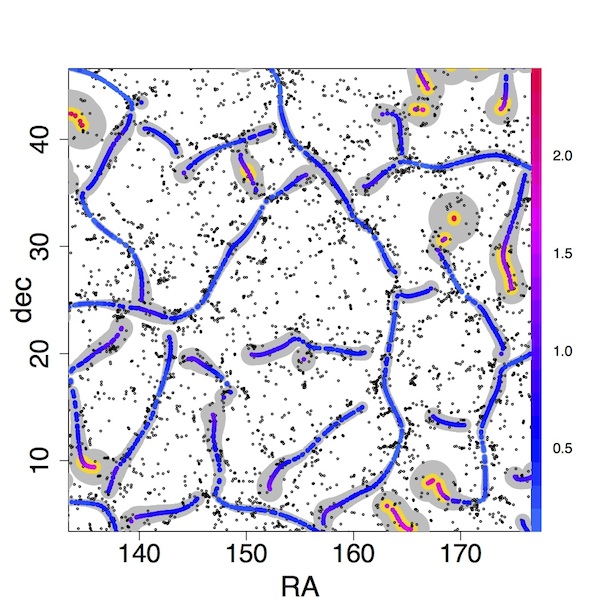}
	}
\subfigure[The smooth bootstrap estimate.]
	{
	\includegraphics[width=2.5 in, height= 2.5 in]{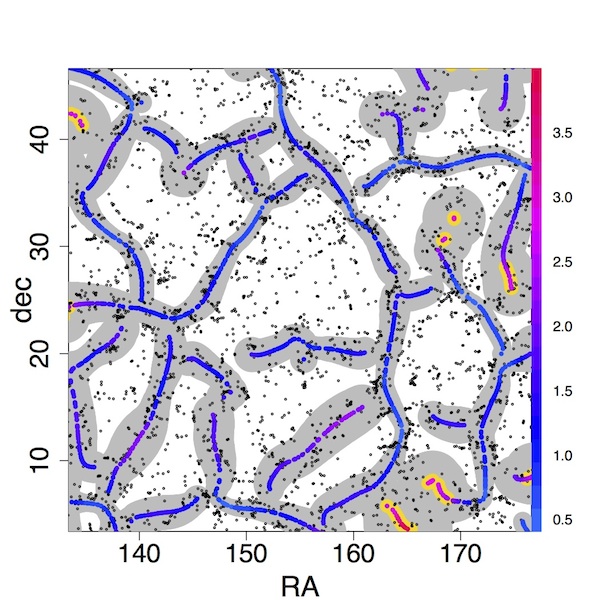}
	}
\caption{Local uncertainty estimates for our high-$z$ SDSS dataset ($z=0.530-0.535$). 
We display the amount of uncertainty via color (red: high) and a confidence band in (a),(b) 
using both ordinary bootstrap and the smooth bootstrap. 
The filament points surrounded by yellow colors are those with high uncertainty
and are declared as `unstable'.
Based on the simulation result in Figure~\ref{fig::FCov}, we expect that the gray regions
in plot (a) contain about $50\%$ true filaments and in (b) contain $85\%$ true filaments.
The unit to the color in uncertainty band is degree.}
\label{fig::LU2}
\end{figure*}

We derive the uncertainties for the filament estimators as described in Section~\ref{sec::LU}
from the two test datasets;
the results for low-$z$ and high-$z$ samples are given in Figures~\ref{fig::LU1}
and \ref{fig::LU2}, respectively.
We visualize local uncertainty using color, where red indicates locations where
the filamentary structure is highly uncertain.
We
also display uncertainty regions as bands of varying width (shown
in gray) centered on the filaments. Our simulation study
in Section \ref{sec::simulation} indicates that the filament coverage
$\FCov$ for the regions in Figures~\ref{fig::LU1}(a) and \ref{fig::LU2}(a)
is $\approx 45\%$, while that in Figures~\ref{fig::LU1}(b) 
and \ref{fig::LU2}(b), is $\approx 60\%$. We find that
the overall structure for filaments in the high-$z$ dataset is more stable
than for the low-$z$ data, due to the significantly larger size of 
the high-$z$ dataset; as shown in Figure~\ref{fig::simN1}, 
sample size plays a crucial role in determining the size of the 
uncertainty regions associated with SCMS filament estimates.

As can be inferred from Figures~\ref{fig::LU1} and \ref{fig::LU2}, our
measures of
local uncertainty provide useful information to determine
the quality of filament detections.
We declare a point $x\in \hat{R}$ 
to be `unstable' if
\begin{equation}
\hat{\rho}(x)\geq \bar{\rho} +1.69\sigma_{\rho},
\label{eq::sig}
\end{equation}
where $\bar{\rho}$ is the mean of uncertainty over all filament points
and $\sigma_\rho$ is the root mean square of uncertainty.
Namely, if the local uncertainty at $x$ is too large,
this point is not stable.
The constant $1.69$ comes from the width of $90\%$ confidence interval
for a Gaussian distribution.
For instance
the two filaments at (RA,$\delta$)$=(165^{\circ}{,}40^{\circ})$ and 
$(165^{\circ}{,}20^{\circ})$ in Figure~\ref{fig::FE1} appear by eye to
be spurious, given the relative lack of galaxies in their vicinity.
Based on the uncertainty measures and our stability test \eqref{eq::sig},
these filaments are declared as unstable (yellow color in Figure~\ref{fig::LU1}).

\subsubsection{Test Data: Comparison to redMaPPer Clusters}	

As one last demonstration of the efficacy of SCMS, we examine the 
consistency between our filament maps and the galaxy clusters
in the redMaPPer catalog 
\citep{2014ApJ...785..104R,2014ApJ...783...80R,2015MNRAS.450..592R}. 
We make this comparison within the window
\begin{align*}
100^{\circ}\leq \mbox{RA}\leq 270^{\circ}~~~-10^{\circ}\leq \delta \leq 70^{\circ}
\end{align*}
and within annuli of width $\Delta z = 0.005$
from $z_{lo}=0.100$ to $z_{hi}=0.500$ (a range that 
includes $10{,}602$ galaxy clusters with spectroscopically determined
redshifts, or $93.1\%$ of the redMaPPer sample).
Note that we also include SDSS DR7 main sample galaxy from NYU VAGC 
\citep{2005AJ....129.2562B,2008ApJ...674.1217P,2008ApJS..175..297A}
to detect filaments for low redshift regions ($z<0.25$).
We slice the data primarily for computational efficiency,
since SCMS is an $O(n^2)$ algorithm, but slicing has the ancillary benefit
of simplifying visualization. In total we examine $80$ slices, each of which
contains $\approx 100$ galaxy clusters. Within each slice, we
determine optimal values of $h$ and $\tau$ using the criteria described 
in Appendix \ref{sec::parameters}.

\begin{figure*}
\centering
\subfigure[$z=0.100-0.150$]
	{
	\includegraphics[scale=0.4]{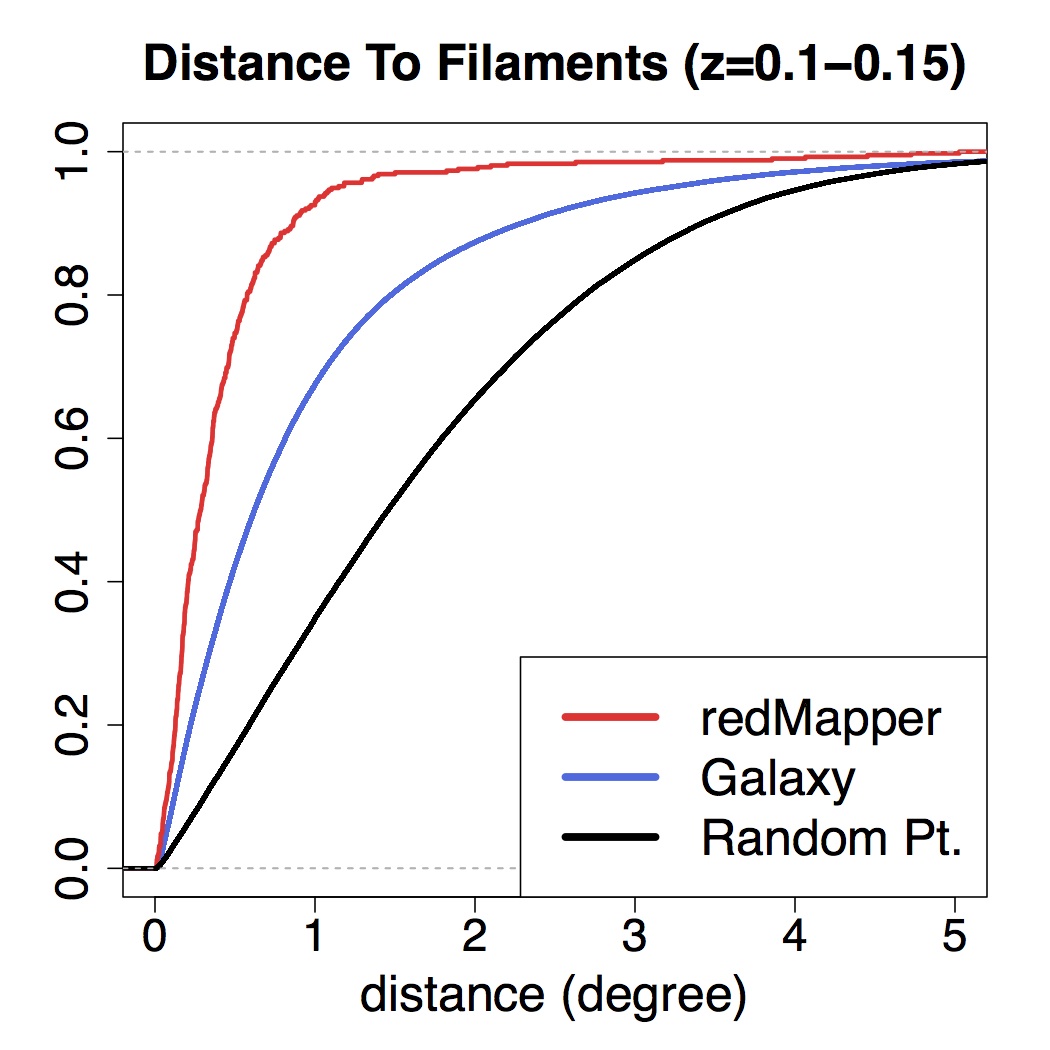}
	}
\subfigure[$z=0.450-0.500$]
	{
	\includegraphics[scale=0.4]{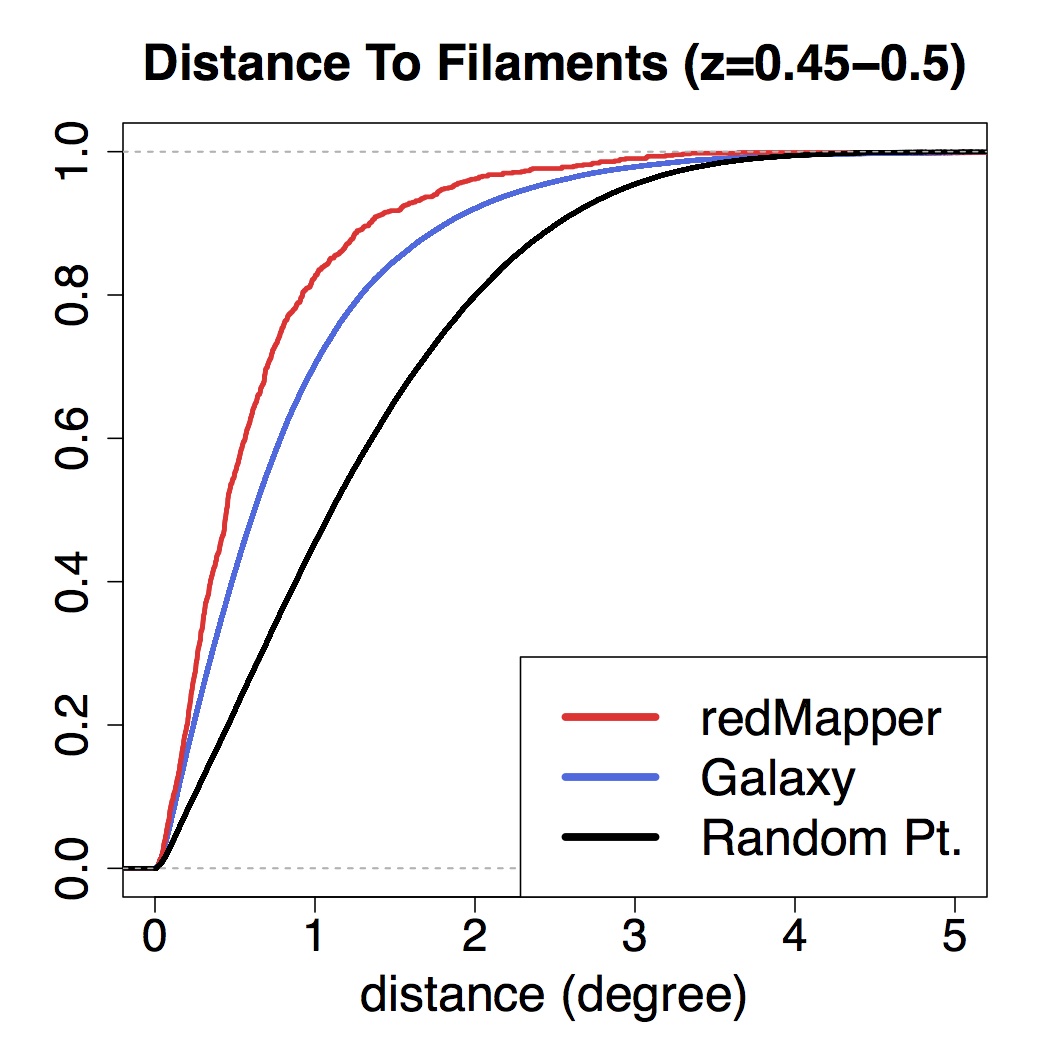}
	}
\caption{The cumulative distribution of the distance statistics 
from galaxies to filaments (blue) 
versus galaxy clusters to filaments (red) at different redshifts.
We also display the distribution for random points (black) as a reference.
The galaxy clusters are from redMaPPer catalog.
The unit of distance is `degree'. 
We only display the first ($z=0.100-0.150$) and the last sub-region ($z=0.450-0.500$)
since other regions have a similar result. The $p$-value for each region is given in Table \ref{tab::RM}.}
\label{fig::DP}
\end{figure*}

In Figure~\ref{fig::BCG}, we display SCMS-detected filaments along
with redMaPPer clusters (in red).
As can be seen, nearly all galaxy clusters are associated
with detected filaments.
Qualitatively similar results hold for all other slices.
To quantify the association of galaxy clusters and filaments,
we compare the distance to filaments for three types of objects:
galaxy clusters, galaxies and randomly generated points 
within the regions where galaxies are observed.
We divide the whole redshift range $z=0.100-0.500$ evenly into $8$ sub-regions
(each sub-region contains 10 slices);
within each sub-region we compute distance statistics.
Ideally, galaxy clusters should be systematically closer to filaments than
galaxies are, and both galaxies and galaxy clusters should be far closer
to filaments than randomly generated points.
Figure \ref{fig::DP} and Table \ref{tab::RM} confirm this hypothesis.
Figure \ref{fig::DP} shows the cumulative distribution for these distance statistics.
For a collection of values $x_1,\cdots,x_n$, the cumulative
distribution function (CDF) is a non-decreasing function ranging from $0$ to $1$
defined as
\begin{equation}
F(x) = \frac{1}{n}\sum_{i=1}^nI(x_i\leq x).
\end{equation}
Both galaxies (blue curves) and galaxy clusters (red curves) 
tend to be much closer to the filaments
than random points; this suggests that galaxies and galaxy clusters
are indeed concentrated around the detected filaments.
When we compare galaxies and clusters,
we observe that galaxy clusters are much more right-skewed in the CDF plot for 
every redshift sub-region.
That is, 
galaxy clusters tend to distribute around low-distance-to-filament regions
compared to a random galaxy.
We conduct the two-sample, one-sided KS test \citep{Stephens1974},
which compares the distributions of distance statistics for galaxy clusters
and randomly generated points, for all eight sub-regions.
Table \ref{tab::RM} shows the $p$-values, a statistical quantity
measuring the significance of observations, for the eight KS tests that we
carry out. A smaller p-value indicates stronger evidence for 
clusters being closer to a filament than galaxies.
Typically, we declare significance as p-value being less than $0.05$.
We observe an increasing trend in $p$-value as the redshift increases, due to
the decrease in the number density of galaxies along the line of sight.
The sharp reversal in this trend at the last sub-region ($z=0.450-0.500$)
is due to the large size of the CMASS sample
at $z>0.430$: the number density of galaxies in our sample
actually increases from $z=0.430-0.500$.

Note that in Figure~\ref{fig::BCG}, many clusters appear to be
located near the intersections of filaments.
However, 
we do not construct a statistics 
to summarize this phenomena since
defining the intersections of filaments detected by
SCMS is a non-trivial problem.
The main difficulty is due to the `gap' between filaments; 
the SCMS filaments will not intersect each other but with a small gap. 
This gap can be explained by the model of density ridges.
In the density ridges model, we require the eigengap $\beta>0$
(recall equation \eqref{eq::eigengap}) 
to ensure the properties of filaments.
Therefore, when one ridge merges with another, 
the eigengap vanishes at some point 
(i.e. $\beta=0$). 
This leaves a small gap between one ridge and another.

\begin{figure*}
\centering
\subfigure[]
	{
	\includegraphics[scale=0.6]{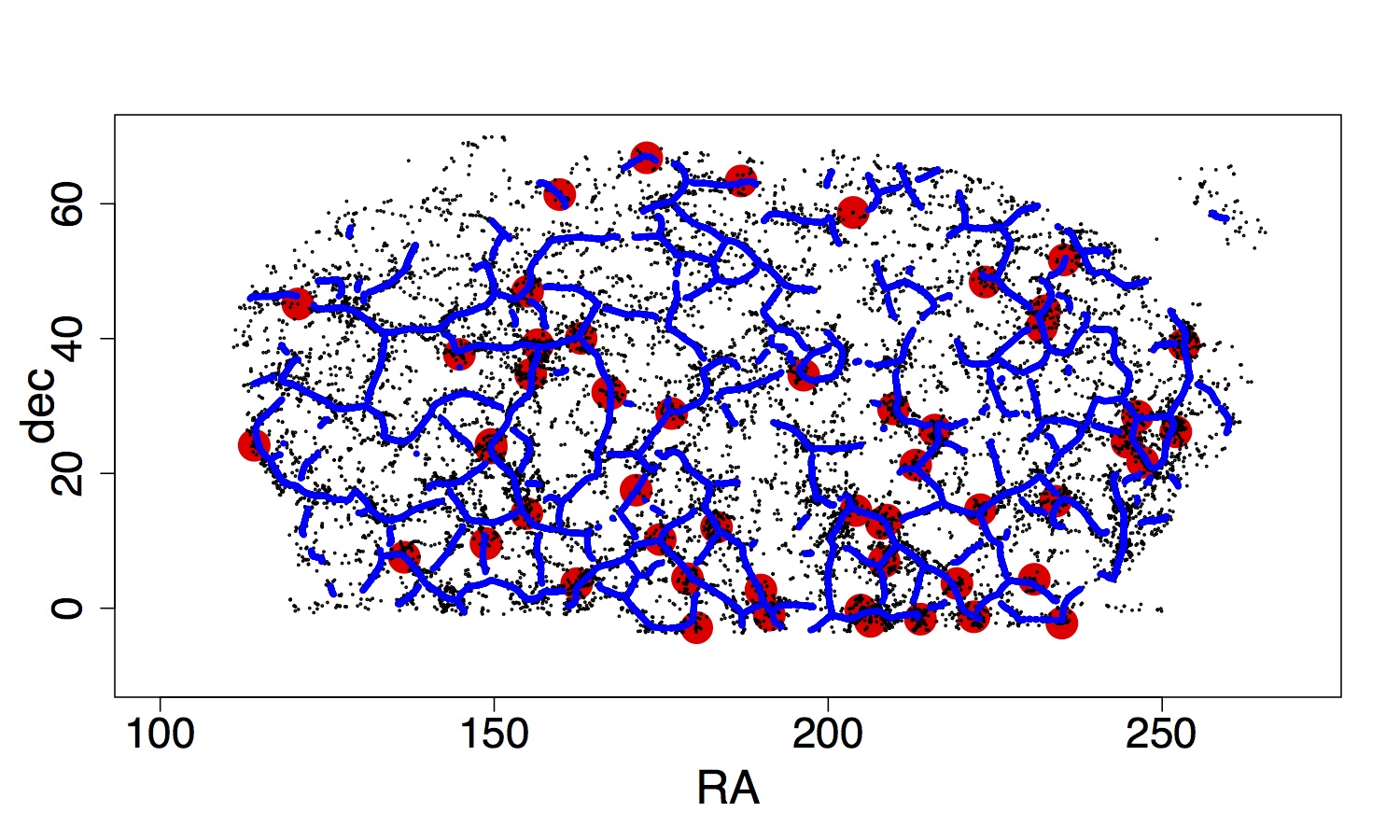}
	}
\caption{Comparison of SCMS filaments to redMaPPer galaxy clusters,
for $z=0.145-0.150$. Shown are SDSS galaxies (black),
putative filaments (blue), and redMaPPer galaxy clusters (red).
As shown in Table \ref{tab::RM}, the redMaPPer clusters
lie significantly closer to filaments than randomly selected points
in the analysis window.}
\label{fig::BCG}
\end{figure*}

\begin{table}
    \begin{tabular}{|l|l|l|l}
    \hline
    Redshift    & $p$-value & Redshift    & $p$-value \\ \hline
    0.100-0.150 & $4.38\times 10^{-40}$       & 0.300-0.350 & $1.73\times 10^{-18}$       \\ \hline
    0.150-0.200 & $1.01\times 10^{-31}$       & 0.350-0.400 & $1.53\times 10^{-10}$       \\ \hline
    0.200-0.250 & $1.66\times 10^{-26}$       & 0.400-0.450 & $7.56\times 10^{-14}$       \\ \hline
    0.250-0.300 & $2.26\times 10^{-19}$       & 0.450-0.500 & $1.95\times 10^{-19}$       \\ \hline
    \end{tabular}
\caption{Significances generated from a one-sided, two-sample KS test, for
the null hypothesis that galaxy clusters lie at the same average distance from
filaments as field galaxies. 
$p$-value is a statistical quantity to measure the significance.
Typically, the usual rejection rule requires $p<0.05$.
The $p$-values show strong evidence that clusters 
lie much closer to filaments than field galaxies.}
\label{tab::RM}
\end{table}

\section{Summary and Discussion}

\label{sec:summary}

In this paper, we demonstrate how one may apply the 
Subspace Constrained Mean Shift (SCMS) algorithm of \cite{Ozertem2011}
to uncover filamentary structure in 
galaxy point cloud data. 
The density ridge model behind the SCMS algorithm ensures
that galaxies will concentrate around detected filaments.
In addition, we introduce an uncertainty measure for detected filaments 
that is based on the bootstrap, allowing us to
study the significance of these filaments.

In {\S}3 we first show that the SCMS filaments are very similar to the Voronoi filaments.
Then we
demonstrate the efficacy of our SCMS-based filament-finding 
algorithm by applying it both to P3M N-body simulation output and to SDSS DR 12
data (including the NYU main sample galaxy, LOWZ and CMASS datasets). 
By applying SCMS to simulated data, we are able to estimate the 
coverage of our bootstrap-generated uncertainty bands, i.e.~the fraction
of any one true filament that lies within its associated band 
(see Figure \ref{fig::FCov}). We find that the coverage depends sensitively
on the number of galaxies in an analyzed sample, with the smooth bootstrap
algorithm generating more conservative uncertainty bands with 1$\sigma$
coverage $\approx$0.6-0.8 (cf.~0.683 for a 1$\sigma$ confidence band) for
galaxy number densities $\approx$5 $\times$ 10$^{-4}$ - 5 $\times$ 10$^{-3}$
(densities observed/to be observed by SDSS CMASS and {\em WFIRST}, 
respectively).

In Figures \ref{fig::FE1}-\ref{fig::LU2}, we show the results of applying
the SCMS algorithm to SDSS spectroscopically observed galaxies in the 
redshift slices 0.235 $\leq z \leq$ 0.240 and 0.530 $\leq z \leq$ 0.535,
respectively. To test the hypothesis that our estimated filaments are 
associated with real filamentary structures, we compare the distances between
filaments and redMaPPer galaxy clusters, random field galaxies, and random 
points in the galaxy field. By using the one-sided, two-sample KS test, we
find that we can safely reject the null hypothesis that galaxy clusters and 
field galaxies reside at similar distances from filaments; the $p$-values
are $\lesssim 10^{-9}$ (cf.~the usual rejection criterion that $p < 0.05$;
see Table \ref{tab::RM}).

The SCMS algorithm models filaments as one-dimensional ridges
that trace high-density regions within the point cloud; as such, SCMS may
be grouped with other filament-modeling algorithms that use the eigenvalues
and eigenvectors of the Hessian matrix associated with the point cloud 
density function, such as MMF \citep{AragonCalvo2007,2010MNRAS.408.2163A} and
NEXUS/NEXUS+ \citep{Cautun2013}.
However, in contrast to these methods,
which output filament estimates as two-dimensional regions,
SCMS filament estimates are smooth, one-dimensional curves; the filament
orientations are well-defined. 
Also in contrast to these methods,
we offer measures of uncertainty by augmenting
the SCMS algorithm with bootstrap-based uncertainty estimation algorithms
that allow one to e.g.~place bands around putative filaments, whose relative
sizes indicate uncertainty in filament location
(as in e.g.~Figure \ref{fig:UMex}). We note that the segmentation-based
DisPerSE algorithm of \cite{2011MNRAS.414..350S} uses
the persistence ratio, a metric encapsulating the evolution of topological 
structure in the galaxy field, to define the {\em significance} of putative 
filaments, but not their spatial uncertainty.
Finally, we compare SCMS filaments to those generated by
the Spine \citep{2010ApJ...723..364A, 2014MNRAS.440L..46A}
and Skeleton \citep{2006MNRAS.366.1201N} algorithms.
Both the Skeleton and Spine models
look for ridges within a density field.
However, the Skeleton model does not provide a means by which to
compute density ridges. In contrast, the SCMS algorithm allows us
to efficiently compute ridges of the field's kernel density estimate.
The Spine method outputs ridges as points on grids,
so that resolution is an issue. On the other hand,
the SCMS algorithm yields points that are on continuous curves (ridges),
so there is no resolution issue to address.


We conclude by stating that one may extend the use of the SCMS algorithm
beyond the analysis of galaxy point cloud data.
For instance, \cite{2014arXiv1406.1803C} discuss how to apply the algorithm to
pixelized image data; in particular, they modify the algorithm
(calling it the weighted SCMS algorithm) to find intensity ridges caused by
e.g.~tidal tails. In addition, the authors also discuss
how one would incorporate the mass of a galaxy to achieve a better estimate
of the local density as well as of corresponding ridges.

\section*{Acknowledgments}

We thank Hy Trac for providing the Nbody simulations and Rachel 
Mandelbaum for useful discussions. 
We also thank the referee Miguel A.~Arag\'on-Calvo for useful comments and 
for providing the Voronoi dataset that we analyze in {\S} \ref{sec::voronoi}.
This work is supported in part by the Department of Energy under 
grant DESC0011114; YC is supported by DOE;
SH is supported in part by DOE-ASC, NASA and NSF; CG
is supported in part by DOE and NSF;
LW is supported by NSF.

Funding for SDSS-III has been provided by the Alfred P. Sloan Foundation, the Participating Institutions, the National Science Foundation, and the U.S. Department of Energy Office of Science. The SDSS-III web site is http://www.sdss3.org/.

SDSS-III is managed by the Astrophysical Research Consortium for the Participating Institutions of the SDSS-III Collaboration including the University of Arizona, the Brazilian Participation Group, Brookhaven National Laboratory, Carnegie Mellon University, University of Florida, the French Participation Group, the German Participation Group, Harvard University, the Instituto de Astrofisica de Canarias, the Michigan State/Notre Dame/JINA Participation Group, Johns Hopkins University, Lawrence Berkeley National Laboratory, Max Planck Institute for Astrophysics, Max Planck Institute for Extraterrestrial Physics, New Mexico State University, New York University, Ohio State University, Pennsylvania State University, University of Portsmouth, Princeton University, the Spanish Participation Group, University of Tokyo, University of Utah, Vanderbilt University, University of Virginia, University of Washington, and Yale University.

\appendix

\section{Parameter Selection}

\label{sec::parameters}

Our version of
SCMS has two key parameters, the smoothing bandwidth $h$ and
the threshold level $\tau$. In this section we show
how we select optimal values for each.

The smoothing bandwidth $h$
represents the amount by which we smooth the observed point cloud of 
galaxies when estimating $p$.
One can choose $h$ by applying prior knowledge or by
letting $h$ adapt to the sample.
There is a large body of literature
on the choice of bandwidth, e.g.
\cite{sheather2004density} and \cite{Chacon2011,chacon2013data}.
Among all methods, we recommend choosing $h$ via
\begin{equation}
h= A_0\times \left(\frac{1}{d+2}\right)^{\frac{1}{d+4}}n^{\frac{-1}{d+4}} \sigma_{\min},
\label{eq::h}
\end{equation}
where $A_0$ is a constant that we discuss below,
$n$ is the sample size, $d$ is the dimension (in our case $d=2$) 
and $\sigma_{\min}$ is the minimal value for the 
standard deviation along each coordinate.
Note that the reference rule \eqref{eq::h} will choose smaller $h$
values as the sample size increases.

\begin{figure*}
\centering
\subfigure[$A_0=1$]
	{
	\includegraphics[width=2.5 in, height= 2.5 in]{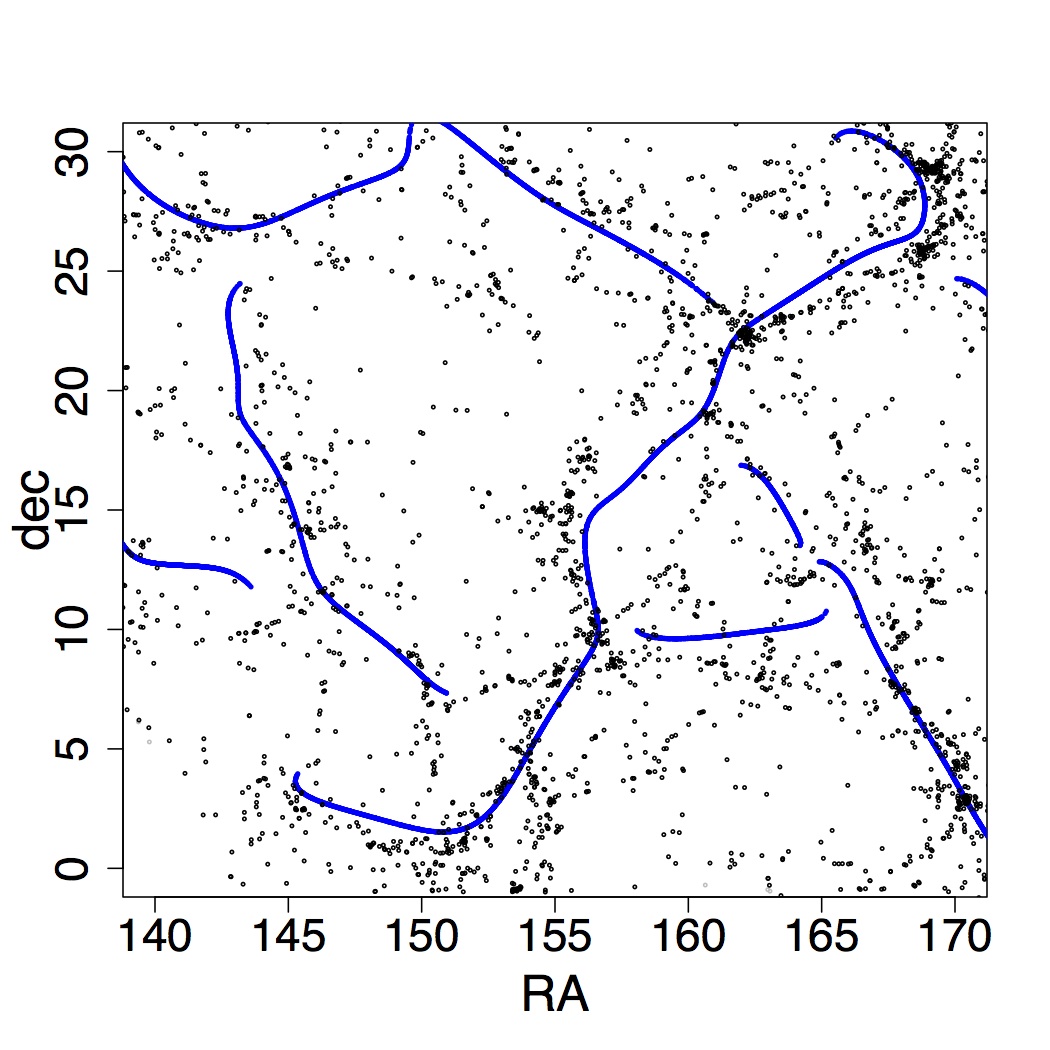}
	}
\subfigure[$A_0=0.6$]
	{
	\includegraphics[width=2.5 in, height= 2.5 in]{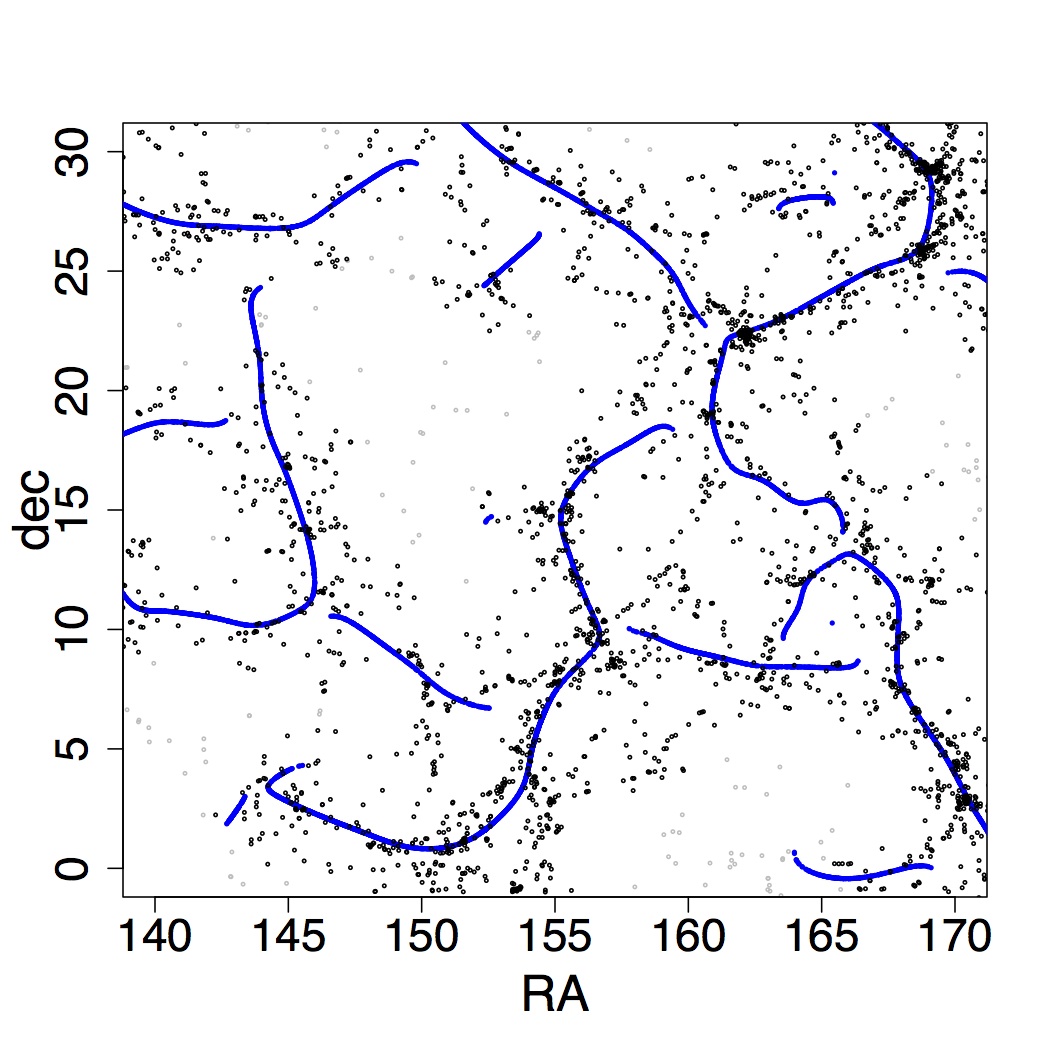}
	}
\subfigure[$A_0=0.2$]
	{
	\includegraphics[width=2.5 in, height= 2.5 in]{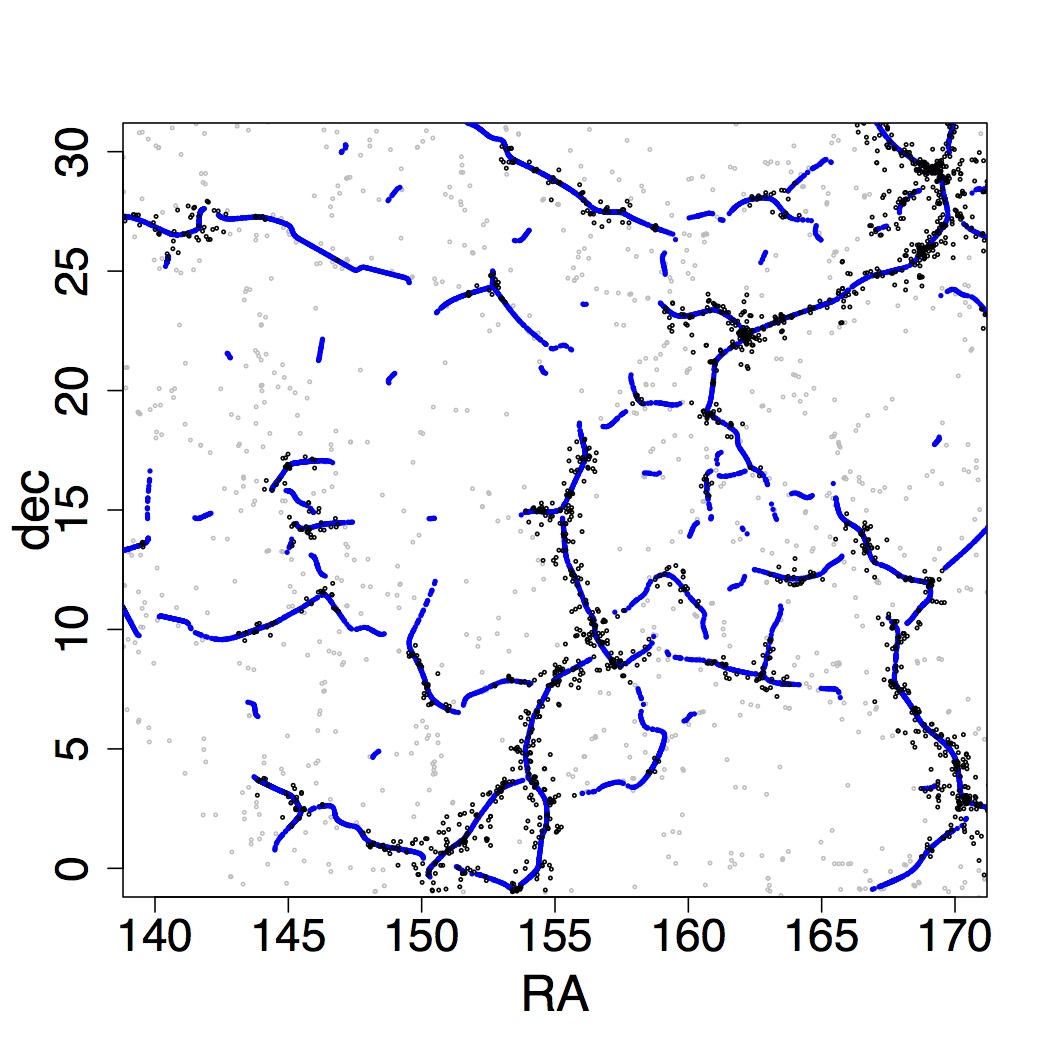}
	}
\subfigure[The comparison.]
	{
	\includegraphics[width=2.5 in, height= 2.5 in]{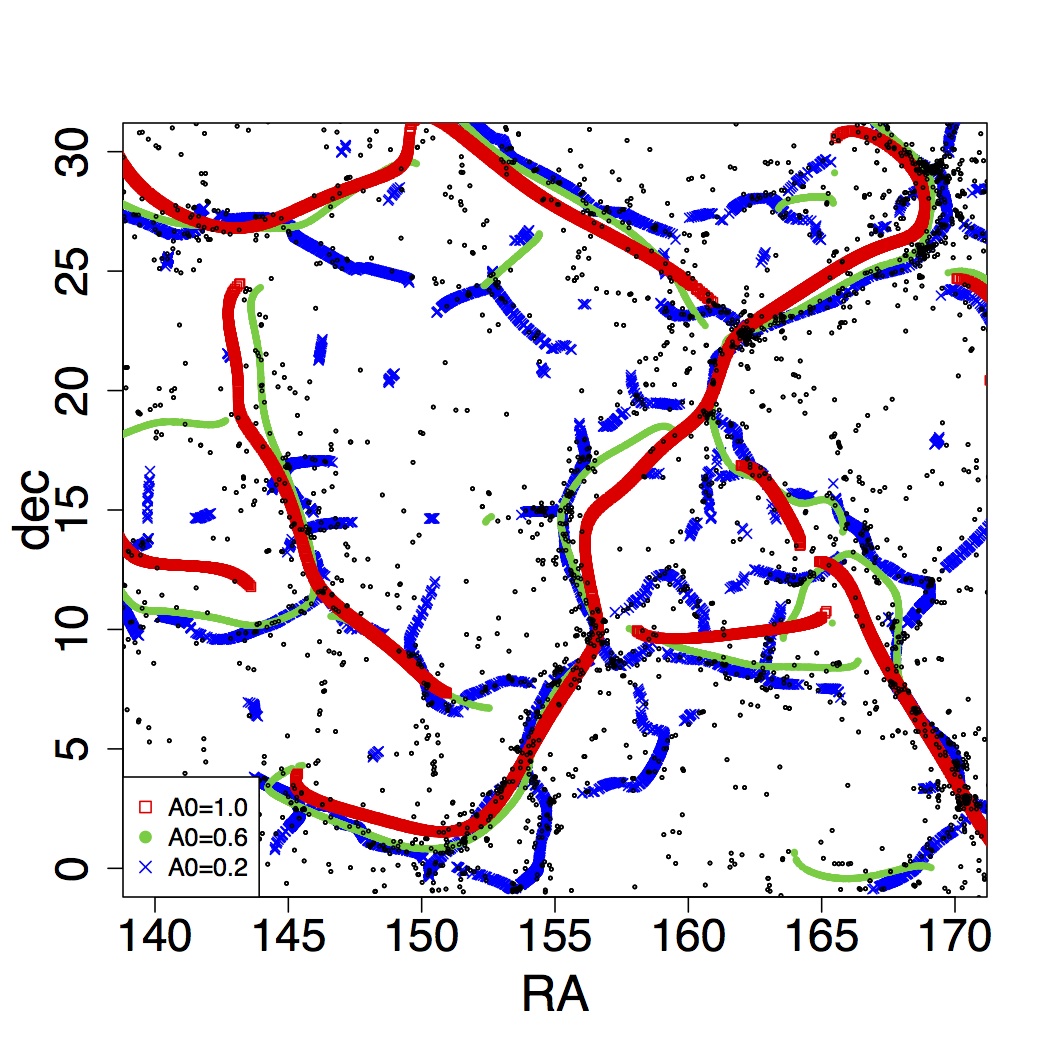}
	}

\caption{The bandwidth selection for equation \eqref{eq::h} for various $A_0$.
The data are galaxies observed spectroscopically by SDSS
in the redshift range $z=0.045-0.050$.
The gray dots are galaxies with density under $\tau  =\sigma( \hat{p})$. 
The black dots are galaxies with density above $\tau$. 
In panel (a)-(c), blue curves are filaments detected by SCMS.
In panel (d), we compare the filaments from (a)-(c).
}
\label{fig::h}
\end{figure*}

\begin{figure*}
\centering
\subfigure[$n=250$, true positive]
	{
	\includegraphics[width=2 in]{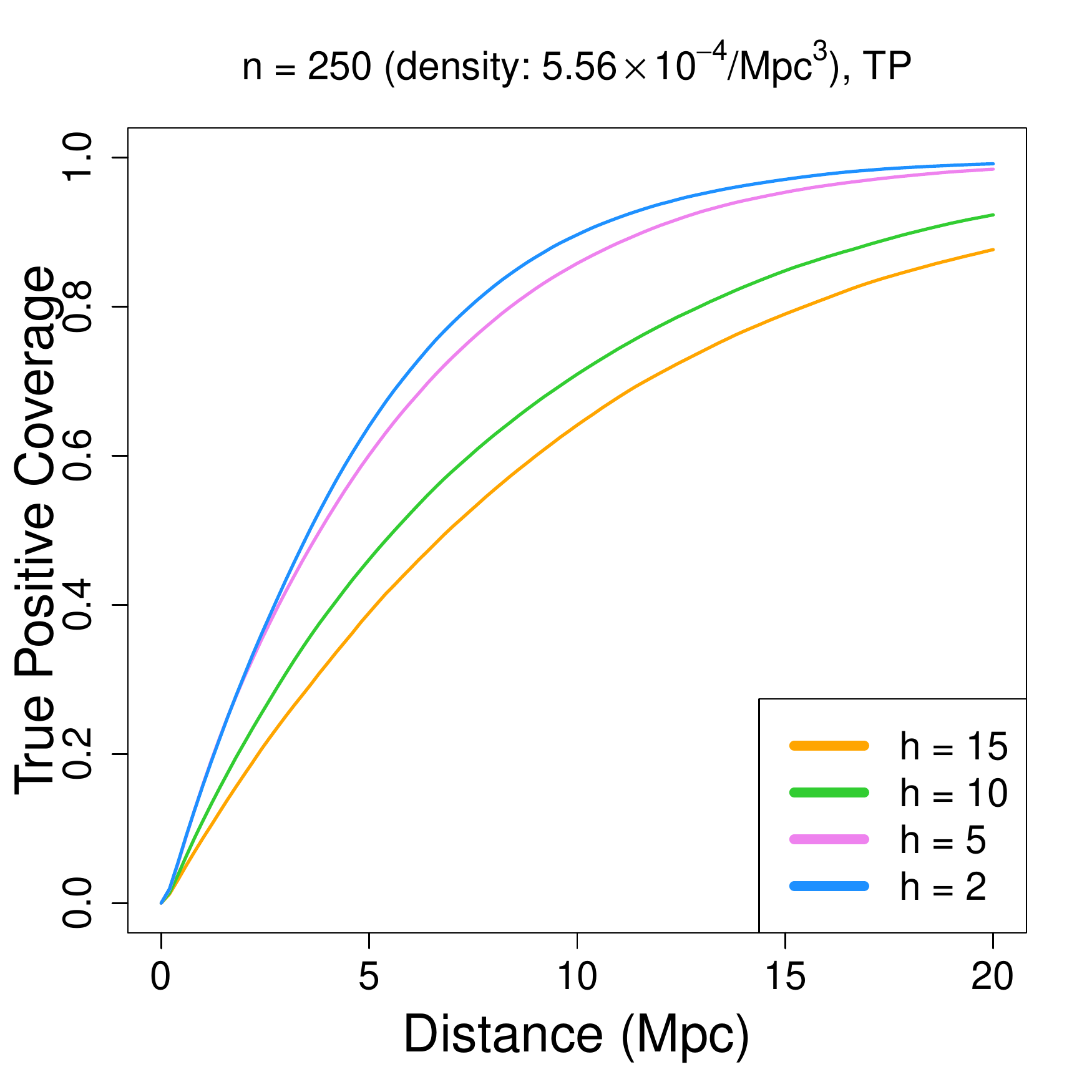}
	}
\subfigure[$n=2500$, true positive]
	{
	\includegraphics[width=2 in]{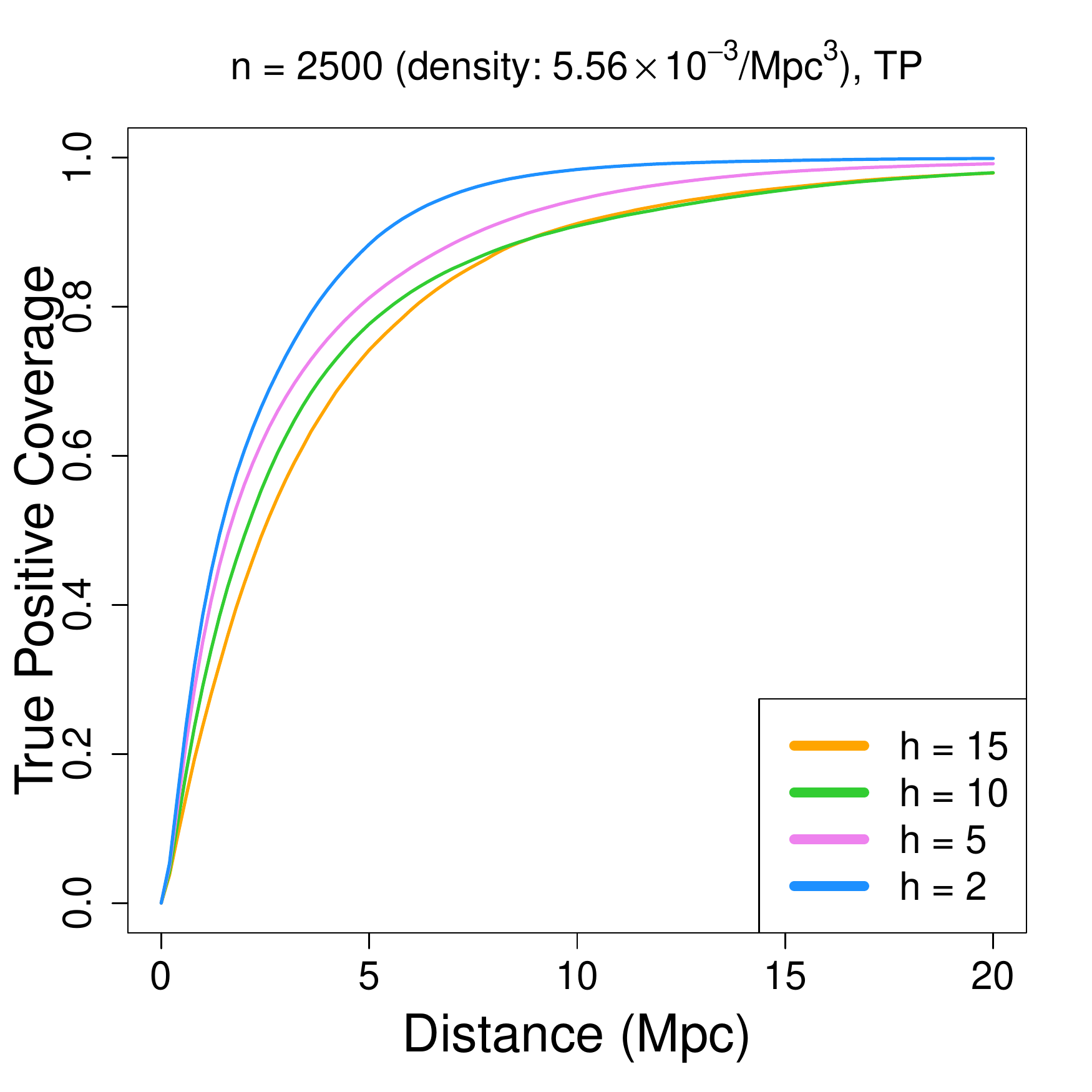}
	}
\subfigure[$n=10000$, true positive]
	{
	\includegraphics[width=2 in]{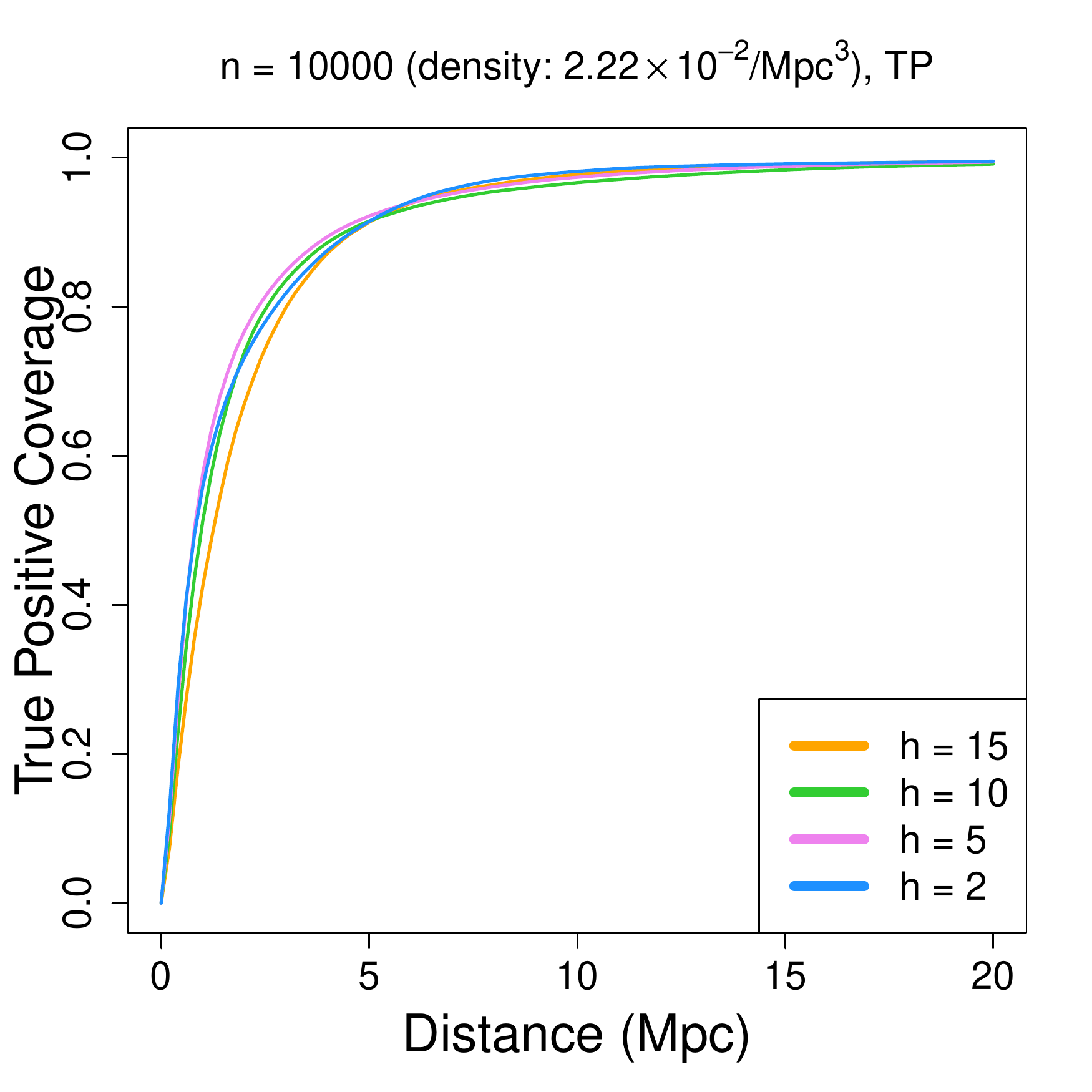}
	}
\subfigure[$n=250$, false positive]
	{
	\includegraphics[width=2 in]{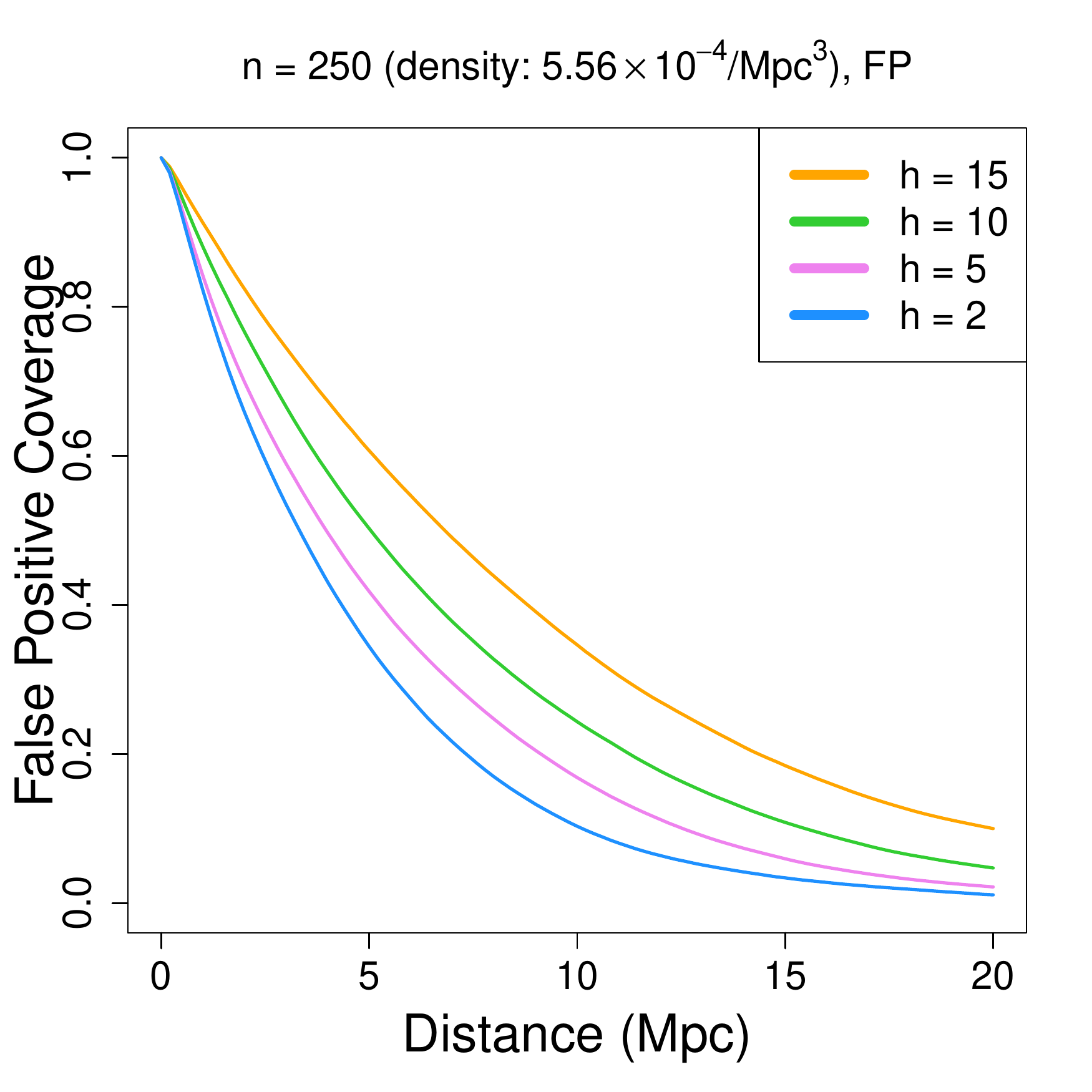}
	}
\subfigure[$n=2500$, false positive]
	{
	\includegraphics[width=2 in]{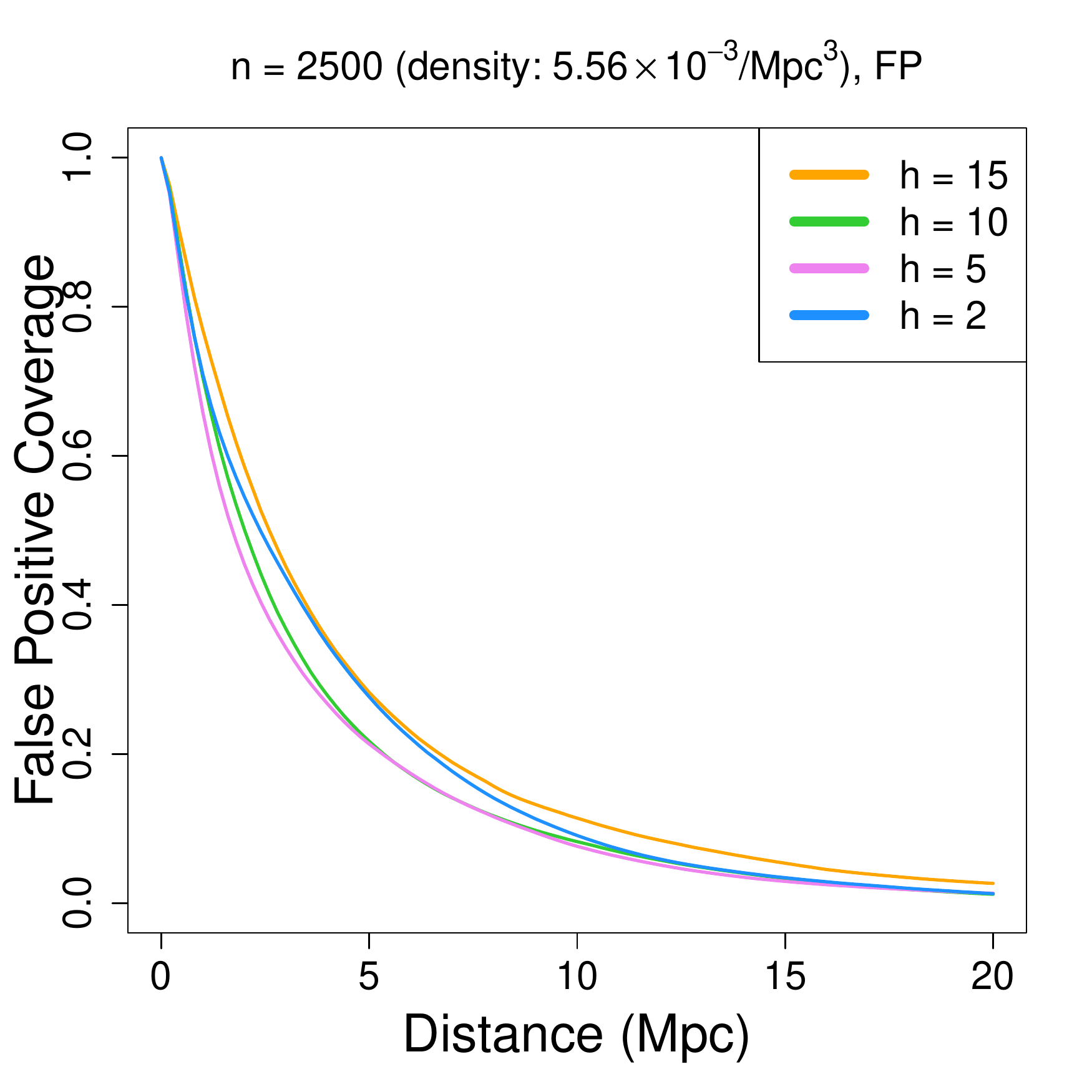}
	}
\subfigure[$n=10000$, false positive]
	{
	\includegraphics[width=2 in]{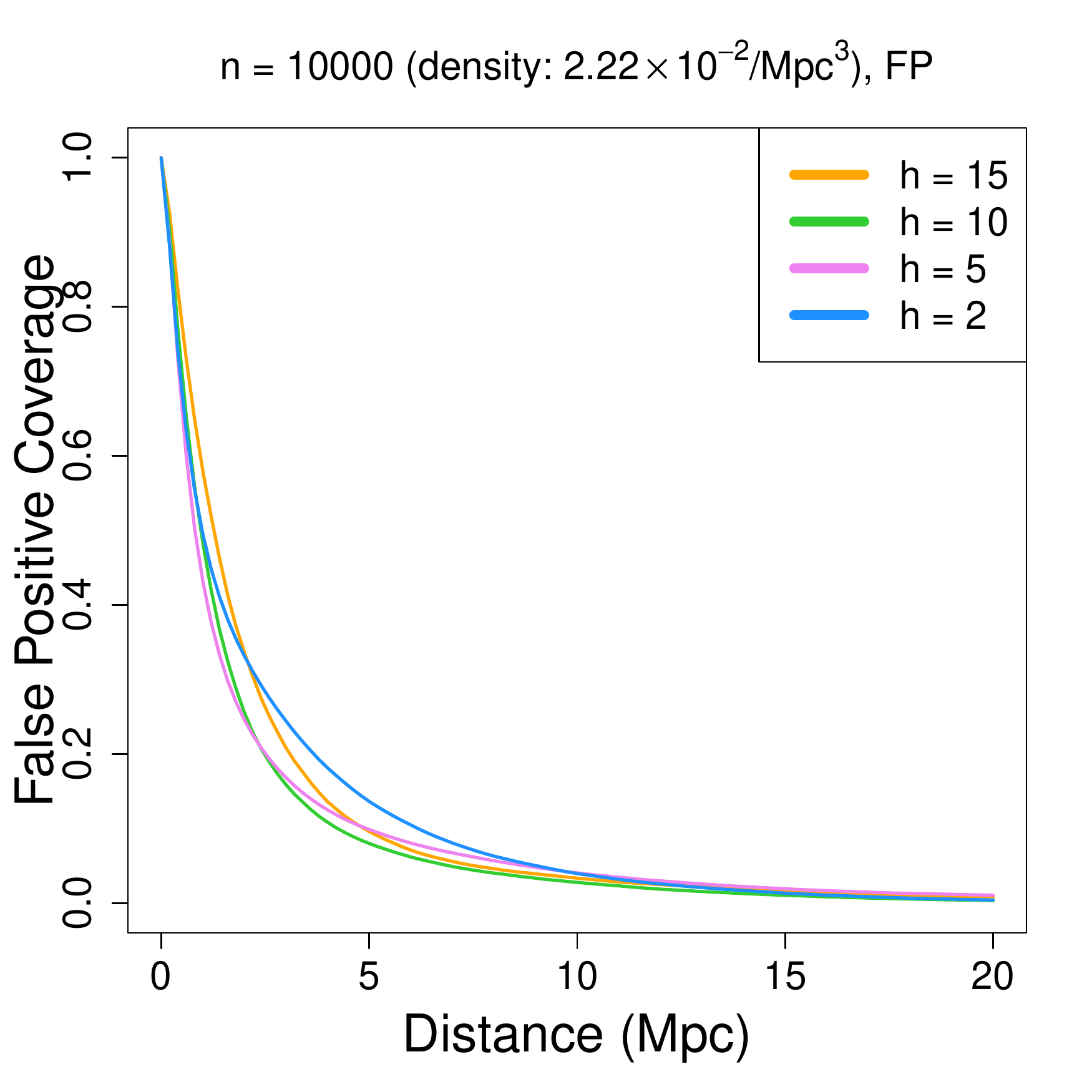}
	}
\caption{The true positive (top row) and false positive (bottom row) coverage
for different sample sizes. As can be seen, $h=5$ or $h=10$ 
(corresponds to the reference rule \eqref{eq::h} using $A_0=0.4$ and $0.8$)
are good choices for both true and false positive coverage.
Note that the reason $h=2$ has good true positive coverage
is because $h=2$ undersmooths the data, leading to
numerous small filaments.
Thus, it is more likely that there are some estimated filaments
around true filaments, which increases the true positive coverage
but also increases the false positive coverage (as true filaments 
may not appear around some estimated filaments).
}
\label{fig::hTPFP}
\end{figure*}

If $A_0$ is $1$, \eqref{eq::h} corresponds to 
Silverman's rule~\citep{Silverman1986}.
Silverman's rule selects $h$ via minimizing the mean integrated error
\begin{equation}
\mathbb{E}\left(\int |\hat{p}(x)-p(x)|^2 dx\right)
\end{equation}
when $p$ is a Gaussian.
When the data include filaments, $p$ is no longer Gaussian
and $A_0$ must be optimized as a free parameter.
A smaller $A_0$ yields more filaments in a given dataset 
but more spurious filaments as well.
There is no general rule for selecting $A_0$ since the optimality criterion
involves the unknown density $p$.
Figure~\ref{fig::h} shows how varying $A_0$
affects the estimation of filamentary structures.
Our results indicate that the optimal $A_0$ lies in the range $[0.4{,}0.8]$.
This is further confirmed by true positive and 
false positive coverage of N-body simulation (described in section~\ref{sec::simulation})
as shown in Figure~\ref{fig::hTPFP}.
In N-body simulation, $A_0=0.4$ corresponds to $h=5 $ Mpc (actual value is $4.82$)
while $A_0 = 0.8$ corresponds to $h=10$ Mpc (actual value is $9.65$).
Both values are better than $h$ being too large or too small 
(compared with $h=2$ Mpc and $h=15$ Mpc cases).
In our analyses of SDSS data, we adopt the value $A_0 = 0.4$.

Thresholding
stabilizes the ridge-finding process since random noise may
cause small bumps in the estimated density field.
However, if the threshold is set too high,
we will remove useful information about the field.
We recommend selecting the thresholding level according to the
root mean square (RMS) in the density fluctuation:
\begin{equation}
\tau = \sigma(\hat{p})\equiv \left(\int_{\mathbb{K}}(\hat{p}(x)-\bar{p}(\mathbb{K}))^2dx\right)^{1/2}\sim \hat{p}-\bar{p},
\label{eq::tau}
\end{equation}
where $\mathbb{K}$ is the region we are interested in
and $\bar{p}(\mathbb{K})$ is the average density in $\mathbb{K}$.
Note that thresholding is also utilized by the
MMF~\citep{2010MNRAS.408.2163A} and NEXUS
\citep{Cautun2013} filament- (and galaxy cluster-) detection algorithms.



\bibliographystyle{mnras}

\bibliography{SuRF.bib}
\bsp

\label{lastpage}

\end{document}